%% file: replica_method_for_nearest_convex_hull_classification.tex
\newcommand{\CLASSINPUTtoptextmargin}{1.7cm}
\newcommand{\CLASSINPUTbottomtextmargin}{1.75cm}
\documentclass[a4paper,conference]{IEEEtran}

\IEEEquantizetextheight{c}
\IEEEsettextwidth{14mm}{14mm}
\IEEEsetsidemargin{c}{0mm}

\usepackage[utf8]{inputenc}
\usepackage[T1]{fontenc}
\usepackage{url}
\usepackage{ifthen}
\usepackage{cite}
\usepackage{subcaption}
\usepackage{bm}
\usepackage[cmex10]{amsmath} 
\usepackage{pgfplots}
\usepackage{amssymb}
\usepackage{tikz, tikzscale}
\pgfplotsset{compat=1.3}
\usetikzlibrary{shapes.geometric,calc,spy}

\DeclareFontFamily{U}{stix2bb}{}
\DeclareFontShape{U}{stix2bb}{m}{n} {<-> stix2-mathbb}{}

\NewDocumentCommand{\indicator}{}{\text{\usefont{U}{stix2bb}{m}{n}1}}

\newtheorem{proposition}{Proposition}
\newtheorem{claim}{Claim}

\newcommand{\squeezeup}{\vspace{-3mm}}

\begin{document}
\tikzset{new spy style/.style={spy scope={%
 magnification=5,
 size=1.25cm, 
 connect spies,
 every spy on node/.style={
   rectangle,
   draw,
   },
 every spy in node/.style={
   draw,
   rectangle,
   fill=gray!5,
   }
  }
 }
} 
\title{Geometric Analysis of Blind User Identification for Massive MIMO Networks}

\author{
    \IEEEauthorblockN{Levi Bohnacker, Ralf R. Müller}
    \IEEEauthorblockA{FAU Erlangen-Nürnberg \\
        Institute for Digital Communications \\
        D-91058 Erlangen, Germany \\
        emails: \{levi.bohnacker, ralf.r.mueller\}@fau.de}
}

\maketitle

\begin{abstract}
    Applying Nearest Convex Hull Classification (NCHC) to blind user identification in a massive Multiple Input Multiple Output (MIMO) communications system is proposed. 
    The method is blind in the way that the Base Station (BS) only requires a training sequence containing unknown data symbols obtained from the user without further knowledge on the channel, modulation, coding or even noise power. 
    We evaluate the algorithm under the assumption of gaussian transmit signals using the non-rigorous replica method. 
    To facilitate the computations the existence of an Operator Valued Free Fourier Transform is postulated, which is verified by Monte Carlo simulation. 
    The replica computations are conducted in the large but finite system by applying saddle-point integration with inverse temperature $\beta$ as the large parameter. 
    The classifier accuracy is estimated by gaussian approximation through moment-matching.
\end{abstract}

\section{Introduction}
Future communication networks are envisioned to enable a multitude of applications through one physical network. 
First realizations of such concepts are observed in 5G Network Slices (NS) which enable to partition the network into virtual sub-networks providing low-latency or high data rates \cite{zhang_overview_2019}. Application oriented slicing is envisioned for 6G, e.g. in the context of Artificial Intelligence (AI) applications \cite{wu_ai-native_2022}. 
At the same time, MIMO is a key enabler of the ever growing demand for high data rates, user density and reliability. 
\\
In this work, a massive MIMO network in which two proprietary applications ($a$ and $b$) share one radio frequency (RF)-chain/BS is considered. 
We assume that the users transmit to the base station at random through a multiple-access scheme mitigating collisions on the network, i.e. no two users transmit at the same time. 
The two networking applications do not reveal the applied modulation and coding scheme or observed Channel-State-Information (CSI) and Signal-to-Noise Ratio (SNR) to the BS.
To forward the received signal to the correct application without leaking information intend for application $a$ to $b$, or vice versa, a blind classifier is applied on the signal received at the BS.
To train the classifier, each user sends a burst of $N$ transmissions along with some application identifier upon synchronization to the BS. 
We observe that this training sequence can be interpreted as noisy samples of a convex hull in high-dimensional space.
Based on this intuition we propose to apply a Nearest Convex Hull Classifier (NCHC) \cite{nalbantov_nearest_2006} for blind user identification. 
The NCHC is computationally expensive in the large dimensional regime and hence subject to theoretical analysis only, from which we can draw insights into the fundamental underlying problem.
Practical implementations may utilize Approximate NCHC (ANCHC) such as the algorithm proposed in \cite{blum_sparse_2019}. 
An empiric comparison of the ANCHC and NCHC in the proposed framework is subject to future work.
\\
Asymptotic analysis of the NCHC is carried out by replica method \cite{mezard_spin_1986}. 
We compare the obtained results with Monte Carlo simulations, showing that the replica computations are precise.
In order to characterize the NCHC, two spin glasses are defined and analyzed in an \textit{isolated} and a \textit{coupled} setting.
The coupling induces a spherical integral which can bee seen as a modified operator-valued version of the famous Harish-Chandra-Itzykson-Zuber (HCIZ) integral \cite{harish-chandra_differential_1957,itzykson_planar_1993}.
To solve the integral asymptotically, a generalization of the Free Fourier Transform \cite{guionnet_fourier_2005} to the operator valued case is postulated, which is to the best of the authors knowledge a novel result. 
An additional modification of the (operator valued) HCIZ integral and the corresponding asymptotics enables the averaging operations during replica computation.

\section{Problem Formulation}
    \subsection{System Model}
\begin{figure}[t]
        \centering
        \def\svgwidth{.5\textwidth} 
        \begingroup%
          \makeatletter%
          \providecommand\color[2][]{%
            \errmessage{(Inkscape) Color is used for the text in Inkscape, but the package 'color.sty' is not loaded}%
            \renewcommand\color[2][]{}%
          }%
          \providecommand\transparent[1]{%
            \errmessage{(Inkscape) Transparency is used (non-zero) for the text in Inkscape, but the package 'transparent.sty' is not loaded}%
            \renewcommand\transparent[1]{}%
          }%
          \providecommand\rotatebox[2]{#2}%
          \newcommand*\fsize{\dimexpr\f@size pt\relax}%
          \newcommand*\lineheight[1]{\fontsize{\fsize}{#1\fsize}\selectfont}%
          \ifx\svgwidth\undefined%
            \setlength{\unitlength}{218.97166335bp}%
            \ifx\svgscale\undefined%
              \relax%
            \else%
              \setlength{\unitlength}{\unitlength * \real{\svgscale}}%
            \fi%
          \else%
            \setlength{\unitlength}{\svgwidth}%
          \fi%
          \global\let\svgwidth\undefined%
          \global\let\svgscale\undefined%
          \makeatother%
        \centering
          \begin{picture}(.75,0.36)%
            \lineheight{1}%
            \setlength\tabcolsep{0pt}%
            \put(0,0){\includegraphics[width=.75\unitlength,page=1]{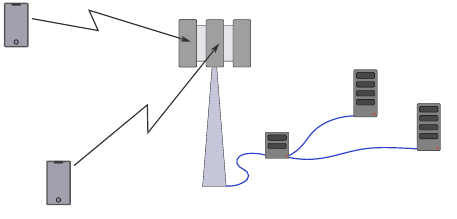}}%
            \put(-0.00105889,0.24156745){\makebox(0,0)[lt]{\lineheight{1.25}\smash{\begin{tabular}[t]{l}$\mathrm{UE}_1$\end{tabular}}}}%
            \put(0.16772318,0.34701661){\rotatebox{-16.91092}{\makebox(0,0)[lt]{\lineheight{1.25}\smash{\begin{tabular}[t]{l}$\mathbf H_a$\end{tabular}}}}}%
            \put(0.18237258,0.15277641){\rotatebox{40.001547}{\makebox(0,0)[lt]{\lineheight{1.25}\smash{\begin{tabular}[t]{l}$\mathbf H_b$\end{tabular}}}}}%
            \put(0.07008116,-0.01756909){\makebox(0,0)[lt]{\lineheight{1.25}\smash{\begin{tabular}[t]{l}$\mathrm{UE}_2$\end{tabular}}}}%
            \put(0.41456409,0.05726836){\makebox(0,0)[lt]{\lineheight{1.25}\smash{\begin{tabular}[t]{l}$\mathrm{CPU}$\end{tabular}}}}%
            \put(0.32906302,0.01139001){\makebox(0,0)[lt]{\lineheight{1.25}\smash{\begin{tabular}[t]{l}$\mathrm{BS}$\end{tabular}}}}%
            \put(0.54484415,0.12632386){\makebox(0,0)[lt]{\lineheight{1.25}\smash{\begin{tabular}[t]{l}$\mathrm{ASPU}_1$\end{tabular}}}}%
            \put(0.65005471,0.07130548){\makebox(0,0)[lt]{\lineheight{1.25}\smash{\begin{tabular}[t]{l}$\mathrm{ASPU}_2$\end{tabular}}}}%
          \end{picture}%
        \endgroup%
        \caption{System architecture: two users communicating through one BS to user-specific, secure applications.}
        \label{fig:system_model}
        \squeezeup
    \end{figure}
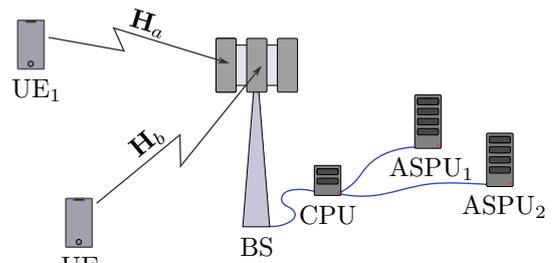

        Consider a communication system comprised of two user equipments (UEs): $\mathrm{UE}_k~\forall~k\in\{a,b\}$; and one BS. 
        The UEs as well as the BS  are equipped with $M$ antennas each.
        The BS RF-chain feeds into a Central Processing  Unit (CPU) which handles the user classification and maps the received signal to the corresponding Application Specific Processing Unit (ASPU) \textit{without} demodulation, as shown in Fig. \ref{fig:system_model}. 
        Upon initial setup and synchronization, $\mathrm{UE}_k$ transmits a burst of $N$ vector-valued symbols $\mathbf x_i\in \mathbb R^{M\times 1}\forall i\in\{1,\ldots,N\}$. 
        The corresponding received signal at the BS is given by:
        \begin{equation}
            \mathbf Y_k = \frac{1}{\sqrt{M}}\mathbf H_k\mathbf X_k + \sigma\mathbf N_k
            \label{eq:transmission_training_sequence}
        \end{equation}
        where $\mathbf H_k=\left(h_{k,i,j}\right)_{i,j=1}^M$ models the channel, $\mathbf N_k = \left(n_{k,i,j}\right)_{i,j=1}^{M,N}$ models \textit{independent identical distributed} (\textit{i.i.d.}) Additive White Gaussian Noise (AWGN) of zero mean and unit variance: $n_{k,i,j}\sim\mathcal{N}(0,1)$ and $\mathbf X_k = [\mathbf x_{k,1},\ldots,\mathbf x_{k,N}] \in\mathbb R^{M\times N}$ is the matrix-valued transmit signal.
        The channel as well as the transmit symbols are modeled by \textit{i.i.d} gaussian coefficients of zero mean and unit variance: $h_{k,i,j}\sim\mathcal N(0,1)$, $\mathbf x_{k,i}\sim\mathcal N(0,\mathbb I_M)$ where $\mathbb I_M\in\mathbb R^{M\times M}$ is the $M\times M$ identity matrix.
        \\
        Upon reception of the training sequences $\mathbf Y_a$ and $\mathbf Y_b$ the BS knows that there are two users in the system associated to application $a$ and $b$. The BS does not have any knowledge of the noise power, signal power or even the underlying system model in \eqref{eq:transmission_training_sequence}.
        Given an unseen received signal $\mathbf y_0$ \eqref{eq:test_transmission}, the BS classifies the application/user based on the received training sequences only.
        \begin{equation}
            \mathbf y_0 = \frac{1}{\sqrt{N}}\mathbf H_l\mathbf x_0 +\sigma\mathbf n_0
            \label{eq:test_transmission}
        \end{equation}
        In the following we give intuition for why the training sequence forms a convex hull in $M$-dimensional space.

    \subsection{An Intuition for Convex Hulls}
\begin{figure*}[t]
        \centering
            \begin{subfigure}[t]{0.3\linewidth}
                \centering
                \begingroup%
                  \makeatletter%
                  \providecommand\color[2][]{%
                    \errmessage{(Inkscape) Color is used for the text in Inkscape, but the package 'color.sty' is not loaded}%
                    \renewcommand\color[2][]{}%
                  }%
                  \providecommand\transparent[1]{%
                    \errmessage{(Inkscape) Transparency is used (non-zero) for the text in Inkscape, but the package 'transparent.sty' is not loaded}%
                    \renewcommand\transparent[1]{}%
                  }%
                  \providecommand\rotatebox[2]{#2}%
                  \newcommand*\fsize{\dimexpr\f@size pt\relax}%
                  \newcommand*\lineheight[1]{\fontsize{\fsize}{#1\fsize}\selectfont}%
                  \ifx\svgwidth\undefined%
                    \setlength{\unitlength}{181.19464424bp}%
                    \ifx\svgscale\undefined%
                      \relax%
                    \else%
                      \setlength{\unitlength}{\unitlength * \real{\svgscale}}%
                    \fi%
                  \else%
                    \setlength{\unitlength}{\svgwidth}%
                  \fi%
                  \global\let\svgwidth\undefined%
                  \global\let\svgscale\undefined%
                  \makeatother%
                  \begin{picture}(.6,.4)%
                    \setlength\tabcolsep{0pt}%
                    \put(0,0){\includegraphics[width=.75\linewidth,page=1]{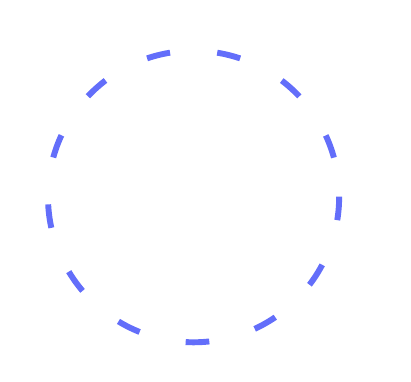}}%
                    \put(0.08,0.56){\makebox(0,0)[lt]{\lineheight{1.25}\smash{\begin{tabular}[t]{l}$\frac{1}{\sqrt{M}}\mathbf X_k$\end{tabular}}}}%
                  \end{picture}%
                \endgroup%
                \caption{Gaussian training sequence}
            \end{subfigure}
            \hfill
            \begin{subfigure}[t]{0.3\linewidth}
                \centering
                \begingroup%
                  \makeatletter%
                  \providecommand\color[2][]{%
                    \errmessage{(Inkscape) Color is used for the text in Inkscape, but the package 'color.sty' is not loaded}%
                    \renewcommand\color[2][]{}%
                  }%
                  \providecommand\transparent[1]{%
                    \errmessage{(Inkscape) Transparency is used (non-zero) for the text in Inkscape, but the package 'transparent.sty' is not loaded}%
                    \renewcommand\transparent[1]{}%
                  }%
                  \providecommand\rotatebox[2]{#2}%
                  \newcommand*\fsize{\dimexpr\f@size pt\relax}%
                  \newcommand*\lineheight[1]{\fontsize{\fsize}{#1\fsize}\selectfont}%
                  \ifx\svgwidth\undefined%
                    \setlength{\unitlength}{196.64584423bp}%
                    \ifx\svgscale\undefined%
                      \relax%
                    \else%
                      \setlength{\unitlength}{\unitlength * \real{\svgscale}}%
                    \fi%
                  \else%
                    \setlength{\unitlength}{\svgwidth}%
                  \fi%
                  \global\let\svgwidth\undefined%
                  \global\let\svgscale\undefined%
                  \makeatother%
                  \begin{picture}(.6,.4)%
                    \setlength\tabcolsep{0pt}%
                    \put(0,0){\includegraphics[width=.75\linewidth]{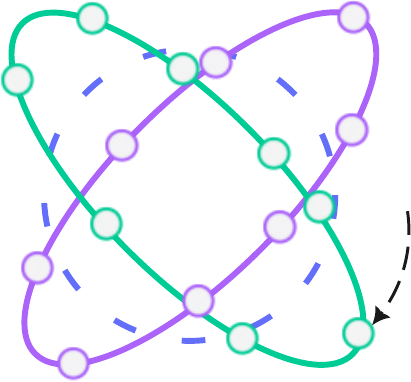}}%
                    \put(.56,0.28){\makebox(0,0)[lt]{\lineheight{1.25}\smash{\begin{tabular}[t]{l}$\frac{1}{M}\mathbf H_a \mathbf x_{a,i}$\\
                    ~~~$+\frac{\sigma}{\sqrt{M}}\mathbf n_{a,i}$\end{tabular}}}}%
                  \end{picture}%
                \endgroup%
                \caption{Channel and noise effect on finite training sequence}
            \end{subfigure}
            \hfill
            \begin{subfigure}[t]{0.3\linewidth}
                    \centering
                    \begingroup%
                      \makeatletter%
                      \providecommand\color[2][]{%
                        \errmessage{(Inkscape) Color is used for the text in Inkscape, but the package 'color.sty' is not loaded}%
                        \renewcommand\color[2][]{}%
                      }%
                      \providecommand\transparent[1]{%
                        \errmessage{(Inkscape) Transparency is used (non-zero) for the text in Inkscape, but the package 'transparent.sty' is not loaded}%
                        \renewcommand\transparent[1]{}%
                      }%
                      \providecommand\rotatebox[2]{#2}%
                      \newcommand*\fsize{\dimexpr\f@size pt\relax}%
                      \newcommand*\lineheight[1]{\fontsize{\fsize}{#1\fsize}\selectfont}%
                      \ifx\svgwidth\undefined%
                        \setlength{\unitlength}{256.78435558bp}%
                        \ifx\svgscale\undefined%
                          \relax%
                        \else%
                          \setlength{\unitlength}{\unitlength * \real{\svgscale}}%
                        \fi%
                      \else%
                        \setlength{\unitlength}{\svgwidth}%
                      \fi%
                      \global\let\svgwidth\undefined%
                      \global\let\svgscale\undefined%
                      \makeatother%
                      \begin{picture}(.5,.45)%
                        \setlength\tabcolsep{0pt}%
                        \put(0,0){\includegraphics[width=.75\linewidth]{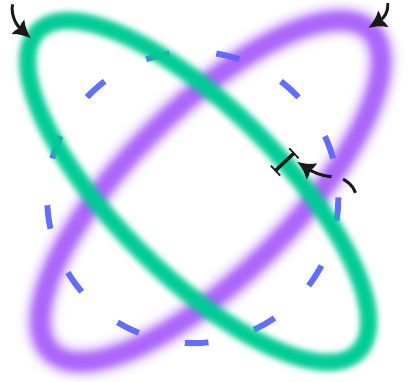}}%
                        \put(-.13,.43){\makebox(0,0)[lt]{\lineheight{1.25}\smash{\begin{tabular}[t]{l}$\scriptstyle\lim\limits_{N\uparrow\infty}\frac{1}{M}\mathbf H_a\mathbf X_a+\frac{\sigma}{\sqrt{M}}\mathbf N_a$\end{tabular}}}}%
                        \put(.26,.43){\makebox(0,0)[lt]{\lineheight{1.25}\smash{\begin{tabular}[t]{l}$\scriptstyle\lim\limits_{N\uparrow\infty}\frac{1}{M}\mathbf H_b\mathbf X_b+\frac{\sigma}{\sqrt{M}}\mathbf N_b$\end{tabular}}}}%
                        \put(.39,.18){\makebox(0,0)[lt]{\lineheight{1.25}\smash{\begin{tabular}[t]{l}$\scriptstyle 2\sigma+\mathcal{O}\left(\frac{1}{\sqrt{M}}\right)$\end{tabular}}}}%
                      \end{picture}%
                    \endgroup%
                    \caption{Channel and noise effect on gaussian training sequence at infinite sample size}
                \end{subfigure}
        \caption{Illustration of the geometry of a high-dimensional training sequence normalized with $\frac{1}{\sqrt{M}}$}
        \label{fig:date_generation_illustration}
        \squeezeup
        \end{figure*}
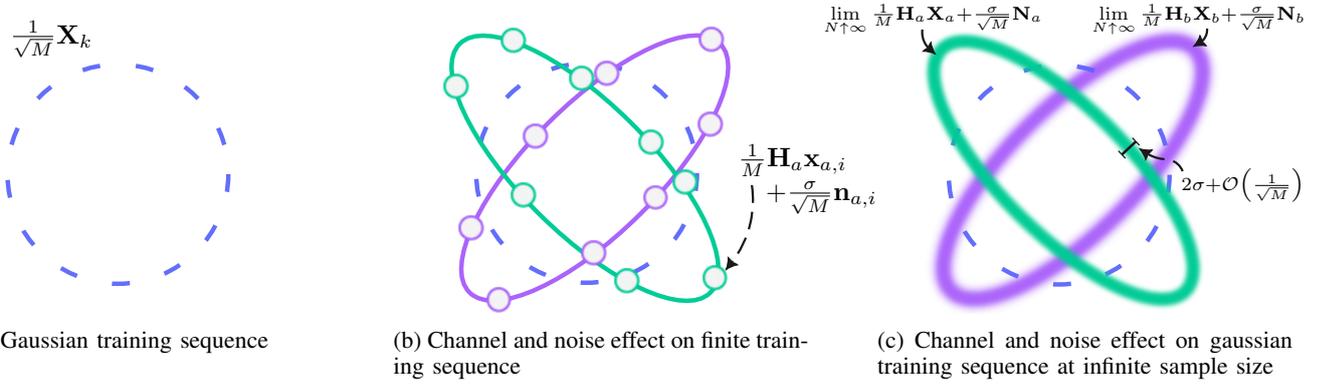

        According to the \textit{Law of Large Numbers} (LLN), the normalized squared $L2$-norm of a high-dimensional \textit{i.i.d.} gaussian vector converges to its variance. It is hence intuitive, that all columns of $\frac{1}{\sqrt{M}}\mathbf X_k$ lie on the $(M-1)$-sphere as illustrated in Fig.\ref{fig:date_generation_illustration} (a).
        Upon transmission through the channel, these gaussian vectors are reshaped in accordance to the singular value distribution of $\frac{1}{\sqrt{M}}\mathbf H_k$ which is described by the Mar\v{c}enko-Pastur Law at $\zeta=1$ \cite{marcenko_distribution_1967}, also known as the Quarter Circle Law \cite{muller_random_2004}. 
        As some singular values are larger than others the hyper-sphere is reshaped to an ellipsoid upon transmission through the channel.
        The AWGN of power $\sigma ^2$ shifts samples of the ellipsoid to lie on $(M-1)$-spheres centered around $\frac{1}{M}\mathbf H_a\mathbf x_{a,i}$, as shown in Fig.\ref{fig:date_generation_illustration} (b).
        Let the training sequence length $N\to\infty$. The additive noise induces a sample distribution centered around the contour of the ellipsoid defined by $\frac{1}{\sqrt{M}}\mathbf H_k$, see Fig.\ref{fig:date_generation_illustration} (c).
        The width of this 'belt' approaches $2 \sigma$ as $M\uparrow\infty$.
        \\
        We interpret the noisy ellipsoid as the typical set of received signals. 
        For sufficiently large $N$, the NCHC compares a new sample to the contour of the ellipsoid, effectively applying typical set decoding.
        This motivates us to analyze the underlying geometry by means of NCHC. 
    
    \subsection{Nearest Convex Hull Classifier}
        The convex hull around a training sequence $\mathbf Y_k$ describes any point $\tilde{\mathbf{y}}$ which can be reached by a convex combination of all samples in $\mathbf Y_k$. The set of all these points is identified by $\mathrm {Conv}(\mathbf Y_k)$:
        \begin{align}
            \mathrm {Conv}(\mathbf Y_k)
            =
            \left\{
                \left.\sum_{i=1}^N \mathbf y_{k,i} v_i \right| ||\mathbf v||_1=1, \mathbf v\in [0,1]^N
            \right\}
        \end{align}
        with $\mathbf Y_k = [\mathbf y_{k,1},\ldots, \mathbf y_{k,N}]$.
        Two training sequences $\mathbf Y_a$ and $\mathbf Y_b$ form two distinct convex hulls. 
        The NCHC classifies a received signal $\mathbf y_0$ by assigning it to the application for which the corresponding convex hull is closer. 
        The squared Euclidean distance of $\mathbf y_0$ to $\mathrm{Conv}(\mathbf Y_k)$ is computed by the following convex optimization problem:
        \begin{align}
            D_{\mathrm l;k} 
            = &\min_{\mathbf v\in[0,1]^N} 
                &\left|\left|
                    \mathbf y_0 - \mathbf Y_k \mathbf v
                \right|\right|_2^2  
            \label{eq:convex_hull_optimization_problem}
                \\&\mathrm{s.t.} &||\mathbf v||_1=1  
            \nonumber     
        \end{align}
        where $l\in\{a,b\}$ identifies the class to which $\mathbf y_0$ belongs. We identify $l=k$ as the Direct Hull (DH) case, i.e. the training sequence and received symbol belong to the same hull and $l\neq k$ as the Cross Hull (CH) case.
        In the binary class case we can define a simple decision metric: $D_l = D_{l;a}-D_{l;b}$. If $D_l<0$, the test point is classified as $a$ and otherwise as $b$. 
        Assuming a uniform prior, we set $l=a$ without loss of generality.
        The classifier accuracy is defined by:
        \begin{align}
            \mathrm{AC}_a(\{ \mathbf Y_k\})
            =~&\mathcal E_{\mathbf y_0|l=a}\left\{\indicator (D_a<0)\right\}
            \\
            =~&\mathrm{Pr}\left(D_a<0 | l=a, \{ \mathbf Y_k\}\right)
        \end{align}
        where $\indicator(\cdot)$ is the indicator function, $\{\mathbf Y_k\}$ identifies $\mathbf Y_a$ and $\mathbf Y_b$ and $\mathcal E_{x|y}\{\cdot\}$ is the expectation operation with respect to (w.r.t.) $x$ given $y$. 
        We aim to compute the average accuracy over all possible channel realizations $\mathbf H_k$, transmit training sequences $\mathbf X_k$, noise realizations $\mathbf N_k$ and received signals $\mathbf y_0$:
        \begin{align}
        \overline{\mathrm{AC}_a}
        =~&\mathcal E_{\mathbf y_0, \{\mathbf Y_k\}|l=a}\left\{\indicator (D_a<0)\right\}
        \label{eq:accuracy_metric_average}
        \end{align}
        where $\overline{X}$ signifies the average of $X$ over all random variables contained in $X$. 

\section{The Replica Computations}
\subsection{Replica Ansatz}
The replica method is a tool from statistical physics which allows to analyze large systems in the thermodynamic equilibrium \cite{mezard_spin_1986}. 
While it is known to miss mathematical rigor, applying the replica method produces results which closely follow numerical simulations and have been shown to be rigorous in certain cases.
We apply the replica method to solve the expectation over the accuracy metric \eqref{eq:accuracy_metric_average}. 
This however is not straightforward as we will be unable to find an analytic continuation around the replica parameter $t$ for $t>0$. 
Instead we leverage Gaussian approximation through moment-matching, computing the mean \eqref{eq:mean_decision_metric} and variance \eqref{eq:variance_decision_metri} of the decision metric by replica method.
\begin{equation}
    \overline{D_a} = \overline{D_{a;a}}-\overline{D_{a;b}}
    \label{eq:mean_decision_metric}
\end{equation}
\begin{align}
    \mathrm{Var}(D_a) =~& \overline{(D_{a;a}-D_{a;b})^2}-\overline{D_{a;a}-D_{a;b}}^2
    \label{eq:variance_decision_metri}
    \\
    =~& 2\mathrm{Var}(D_{a;a}) + 2\mathrm{Var}(D_{a;b}) - \mathrm{Var}(D_{a;a}+D_{a;b})
    \nonumber
\end{align}
The spin glass energy function is defined in a general setting by the Hamiltonian \eqref{eq:general_hamiltonian}.
\begin{align}
    \mathcal H(\mathbf v_\mathrm{a},\mathbf v_\mathrm{b}|\{\mathbf Y_i\},\mathbf y_0)
    =~&
    \tau_{a}\left|\left| 
        \mathbf y_0-\mathbf Y_a\mathbf v_\mathrm{a}
    \right|\right|^2_2
    \label{eq:general_hamiltonian}
    \\
    \nonumber
    &
    + \tau_{b}\left|\left| 
        \mathbf y_0-\mathbf Y_b\mathbf v_\mathrm{b}
    \right|\right|^2_2
\end{align}
Parameters $\tau_a, \tau_b \in \{0,1\}$ allow to combine the DH and CH terms.
Selecting $\tau_a=\tau_b=1$ is referred to as the \textit{coupled} spin glass whereas $\tau_a\neq\tau_b$ yields an \textit{isolated} spin glass.
To reduce notational overhead we identify all random variables by $\Omega$.
The corresponding free energy is given by:
\begin{align}
    F(\beta|\Omega) &= 
    -\frac{1}{\beta M } \log\int
    \mathrm d \mathbf v_a \mathrm d \mathbf v_b
    e^{-\beta \mathcal H(\mathbf v_a,\mathbf v_b|\Omega)}
    \delta_{||\mathbf v_a||_1}\delta_{||\mathbf v_b||_1}
    \label{eq:general_free_energy_hamiltonian}
    \\
    &=-\frac{1}{\beta M }\log Z(\beta|\Omega)
    \label{eq:general_free_energy_partition_function}
\end{align}
The integral is taken on the domains $\mathbf v_a,\mathbf v_b\in[0,1]^{N\times 1}$, $\delta_{||\mathbf v_k||_1}=\delta(||\mathbf v_k||_1-1)$ enforces the unit $L1$-norm constraint on the two microstate variables $\mathbf v_a$ and $\mathbf v_b$, $Z(\beta|\Omega)$ is the partition function and $\beta$ is the inverse temperature parameter.
In order to obtain the statistics of the free energy we define the Scaled Cumulant Generating Function (SCGF) \eqref{eq:SCGF_free_energy_form}.
\begin{align}
    G(\beta,t) &= -\frac{1}{\beta M} \log \overline{Z(\beta|\Omega)^{t}}
    \\
    &= -\frac{1}{\beta M} \log \overline{e^{-t\beta M F(\beta|\Omega)}}
    \label{eq:SCGF_free_energy_form}
\end{align}
The first and second order derivative of \eqref{eq:SCGF_free_energy_form} w.r.t. $t$ at $t=0$ yields the mean and scaled variance of the free energy, respectively:
\begin{equation}
    \left.\frac{\partial}{\partial t} G(\beta,t)\right|_{t=0}
    = \overline{F(\beta|\Omega)}
    \label{eq:SCGF_mean_free_energy}
\end{equation}
\begin{equation}
    \left.\frac{\partial^2}{\partial t^2} G(\beta,t)\right|_{t=0}
    = -\beta M\left(\overline{F(\beta|\Omega)^2}-\overline{F(\beta|\Omega)}^2\right)
    \label{eq:SCGF_var_free_energy}
\end{equation}
As the spin glass freezes: $\beta\uparrow\infty$; the integral in \eqref{eq:general_free_energy_hamiltonian} is solved by saddle-point integration where the extrema of the exponent correspond to the solution of \eqref{eq:convex_hull_optimization_problem}.
Computing the free energy cumulants at inverse temperature $\beta\uparrow\infty$ hence provides the cumulants of the decision metric $D_a$. 
Obtaining a closed form solution to the SCGF $G(\beta,t)$ therefore facilitates the computation of \eqref{eq:SCGF_mean_free_energy} and \eqref{eq:SCGF_var_free_energy}.
The following paragraphs outline the replica computations corresponding to the isolated and coupled spin glass, respectively.

\subsection{The Isolated Spin Glasses}
The CH and DH isolated spin glasses are not identical. 
We hence present a generalized approach parametrized by $k=a$ on $\tau_a=1$ (DH) and $k=b$ on $\tau_b=1$ (CH). 
Explicit analytic function forms are omitted, in order to reduce notational overhead. 
The averaged replicated partition function of the isolated spin glass is denoted by $f_{Z,k}(\beta,t)~\forall~k\in\{a,b\}$ and referred to as the isolated moments function.
\begin{align}
    f_{Z,k}(\beta,t)
    =&
    \int  \mathcal E_{\Omega}\left\{
    e^{-\beta\sum_{r=1}^t \left|\left|\mathbf H_a\mathbf x_0+\sigma \mathbf n_0-\left(\mathbf H_k\mathbf X_k+\mathbf N_k\right)\mathbf v_{k,r}\right|\right|_2^2}
    \right\}
    \nonumber
    \\ &\times
    \prod_{r=1}^t
     \delta_{||\mathbf v_{k,r}||}
     \mathrm d \mathbf v_{k,r}
     \label{eq:moments_function_isolated_spin_glass_initial}
\end{align}
In a first step, the average over $\mathbf N_k$ is computed by gaussian integration. 
Then, we jointly average over $\mathbf x_0$ and $\mathbf n_0$ by interpreting $\mathbf y_0$ as a multivariate Gaussian of zero-mean and covariance $\mathbf C_{\mathbf y_0} = \sigma^2\mathbb I_M+\frac{1}{M}\mathbf H_a\mathbf H_a^T$.
This two-fold gaussian integration yields:
\begin{align}
    f_{Z,k}(\beta,t)
    =~&
    \int  \mathcal E_{X,H}\left\{
    e^{-\beta\alpha N\sum_{r=1}^t \mathrm{tr}\left\{\mathbf D_{k,r} \tilde{\mathbf X}_k^T\mathbf A_{k,r}\tilde{\mathbf X}_k\right\}\
    +\mathcal D(\mathbf \Lambda_k)
    }
    \right\}
    \nonumber
    \\ 
    &\times
    \prod_{r=1}^t
     \delta_{||\mathbf v_{k,r}||}
     \mathrm d \mathbf v_{k,r}
\end{align}
where $\alpha=\frac{M}{N}$, $\mathbf \Lambda_k$ is a diagonal $N\times N$ matrix containing the eigenvalues of $\mathbf V_k\mathbf V_k^T$, $\mathbf D_{k,r}$ is a rank-one diagonal matrix with a non-zero entry $\theta_r$ at index $(r,r)$ which is a function of the $r^{th}$ eigenvalue of $\mathbf V_k\mathbf V_k^T$, $\tilde{\mathbf X}_k=\frac{1}{\sqrt{M}}\mathbf U\mathbf X_k$ and $\mathbf A_{k,r}$ is a function of $\frac{1}{\sqrt{M}}\mathbf H_k$, $\sigma$, $\beta$ and $\mathbf V_k\mathbf V_k^T$.
With $\mathbf V_k = [\mathbf v_{k,1}, \ldots, \mathbf v_{k,t}]\in[0,1]^{N\times t}$.
Furthermore, the function $\mathcal D:\mathbb R^{N\times N}\to\mathbb R$ is induced by the covariance scaling of the  gaussian integral and is self-averaging w.r.t. $\mathbf H_a$.
The empirical singular-value spectrum (ESD) of $\tilde{\mathbf X}_k$ and $\mathbf A_{k,r}$ converges and is bounded.
We hence only have to average over the left unitary matrix of the Singular Value Decomposition (SVD) of $\tilde{\mathbf X}_k$ which is distributed by a Haar measure on the orthogonal group: $\mu(\cdot)$.
\begin{equation}
    \Xi_{X,k}
    =
    \int 
    \exp\left\{-\beta\alpha N\sum_{r=1}^t \theta_r  \mathbf u_{k,r}^T \tilde{\mathbf X}_k^T\mathbf A_{k,r}\tilde{\mathbf X}_k\mathbf u_{k,r}\right\}
    \mathrm d\mu(\mathbf U_k)
    \label{eq:modified_HCIZ_integral}
\end{equation}
with $\mathbf U_k = [\mathbf u_{k,1},\ldots, \mathbf u_{k,N}]$ and $\Xi_{X,k}$ being a symbolism for the expectation over $\mathbf X_k$.
The integral \eqref{eq:modified_HCIZ_integral} is a spherical integral.
In fact, if $\mathbf A_{k,r}=\mathbf A_k ~\forall ~r$ were to hold, this would be a reformulation of the well known HCIZ integral \cite{harish-chandra_differential_1957,itzykson_planar_1993}, which is solved asymptotically in the low-rank regime using the Free Fourier Transform \cite{guionnet_fourier_2005}.
By modifying the derivations in \cite{guionnet_fourier_2005}, we allow the inner matrix $\mathbf A_{k,r}$ to be dependent on the index $r$ such that:
\begin{equation}
    \Xi_{X,k} \doteq e^{
        -\beta \alpha N 
        \sum_{r=1}^t 
            \int_0^1 
                \theta_rR_{\tilde{X}_k^T\mathbf A_{k,r}\tilde{\mathbf X}_k}\left(
                    -2 \beta \alpha \theta_rw
                \right)
                \mathrm d w
    }
    \label{eq:modified_free_fourier_transform}
\end{equation}
where $\doteq$ signifies asymptotic equivalence and $R_{\mathbf B}(\cdot)$ is the R-transform of $\mathbf B$ \cite{speicher_free_2009}.
Considering that $\theta_r$ is a function of the $r^{th}$ eigenvalue of $\mathbf V_k\mathbf V_k^T = \mathbf G(\{\mathbf v_{k,r}\})$, \eqref{eq:modified_free_fourier_transform} is given in matrix notation by \eqref{eq:Xi_X_k_matrix_notation}.
\begin{equation}
    \Xi_{X,k} \doteq e^{
    -\beta \alpha N 
    \int_{0}^1
    \mathrm{tr}\left\{
        \mathbf G(\{\mathbf v_{k,r}\})
        R_{k,\{r\}}\left(
            -2 \beta \alpha w\mathbf G(\{\mathbf v_{k,r}\})
        \right)
    \right\}
    }
    \label{eq:Xi_X_k_matrix_notation}
\end{equation}
The matrix valued function $R_{k,\{r\}}$ applies the R-transform $R_{\tilde{\mathbf X}_k^T\mathbf A_{k,r}\tilde{\mathbf X}_k}(\cdot)$ to the $r^{th}$ eigenvalue of its argument. 
In \eqref{eq:Xi_X_k_matrix_notation}, $\Xi_{X,k}$ is fully characterized by the eigenvalues of $\mathbf G(\{\mathbf v_{k,r}\})\in[0,1]^{N\times N}$.
As these are identical to the eigenvalues of $\mathbf Q(\{\mathbf v_{k,r}\})=\mathbf V_k^T\mathbf V_k \in \mathbb R^{t\times t}$, $\Xi_X$ can be expressed as a function of this \textit{replica correlation} matrix.
To simplify the integral over $\{\mathbf v_{k,r}\}$, the subshell \eqref{eq:subshell_definition} is defined.
\begin{equation}
    \mathbb S(\mathbf Q_k) = 
    \left\{
        (\mathbf v_{k,1},\dots, \mathbf v_{k,t}|\mathbf v_{k,i}^T\mathbf v_{k,j}=q_{k,ij})
    \right\}
    \label{eq:subshell_definition}
\end{equation}
with $\mathbf Q_k=(q_{k,i,j})_{i,j=1}^t$.
By subshell integration of \eqref{eq:moments_function_isolated_spin_glass_initial} after averaging over $\Omega$, the moments function \eqref{eq:moments_function_subshell} is obtained.
\begin{equation}
    f_{Z,k}(\beta,t) 
    =
    \int \mathrm d\mathbf Q_k
    e^{
        -\mathcal K_k(\mathbf Q_k)
        -\mathcal D(\mathbf Q_k)
        +\mathcal I(\mathbf Q_k)
    }
    \label{eq:moments_function_subshell}
\end{equation}
with:
\begin{align}
    \mathcal K_k(\mathbf Q_k)
    =~&\beta \alpha N 
    \int_{0}^1
    \mathrm{tr}\left\{
        \mathbf Q_k
        R_{k,\{r\}}\left(
            -2 \beta \alpha w \mathbf Q_k
        \right)
    \right\}
    \mathrm d w
    \\
    \mathcal D(\mathbf Q_k)
    =~&
    \frac{M}{2}\log\left|
        \mathbb I_t + 2\beta\sigma^2 \mathbf Q_k
    \right|
    +\frac{1}{2}\log \left|
        \sigma^2\mathbb I_M + \mathbf H_a\mathbf H_a^T
    \right|
    \nonumber
    \\ 
    &+\frac{1}{2}\log|(\sigma^2\mathbb I_M + \mathbf H_a\mathbf H_a)^{-1} + e(\mathbf Q_k)\mathbb I_b|
    \label{eq:isolated_determinant_terms}
\end{align}
and the subshell density $\mathcal I(\mathbf Q_k)$:
\begin{align}
    \mathcal I(\mathbf Q_k)
    =~&
    \log
    \int \delta(\mathbf Q-\mathbf V_k^T\mathbf V_k)\prod_{r=1}^t\delta_{||\mathbf v_{k,r}||_1} \mathrm d_{\mathbf v_{k,r}}
    \\
    =~&
    \log
    \int \delta(\beta\mathbf Q-\beta\mathbf V_k^T\mathbf V_k)\prod_{r=1}^t\delta(\beta||\mathbf v_{k,r}||_1-\beta) \mathrm d_{\mathbf v_{k,r}}
    \nonumber
    \\
    &+
    (t+1)\log(\beta)
\end{align}
Averaging over $\mathbf H_a$ is not done explicitly as the log-determinants are self-averaging at large $N$ and $M$.
To compute the sub-shell density we follow the approach taken in \cite{bereyhi_statistical_2019} and obtain \eqref{eq:subshell_integration}. The summand $(t+1) \log(\beta)$ is ignored in subsequent steps as it will vanish for large $\beta$ upon normalization with $\frac{1}{\beta}$ in \eqref{eq:general_free_energy_hamiltonian}.
\begin{align}
    e^{\mathcal I(\mathbf Q_k)}
    = 
    \int
    \mathrm d \mathbf S_k \mathrm d \mathbf r_k 
        e^{
            -\beta \mathrm{tr}\left\{
                \mathbf S_k^T\mathbf Q_k
            \right\}
            -\beta \mathbf r_k^T\mathbf 1_{t\times 1}
            +\mathcal P(\mathbf S_k,\mathbf r_k)
        }
        \label{eq:subshell_integration}
\end{align}
with: 
\begin{align}
    \mathcal P(\mathbf S_k, \mathbf r_k)
    =
    N \log \int_{\mathbf v\in[0,1]^t}
    \exp\left\{
        \beta \mathbf v^T\mathbf S_k \mathbf v + \beta \mathbf r_k^T\mathbf v
    \right\}
    \mathrm d \mathbf v
\end{align}
Replica-Symmetry (RS) parametrization is applied to $\mathbf Q_k,\mathbf S_k$ and $\mathbf r_k$.
\begin{align}
    \mathbf Q_k 
    =~&
    \frac{\chi_k}{\beta}\mathbb I_t + q_k \mathbf1 _{t\times t}
    \\
    \mathbf S_k=~& - e_k \mathbb I_t + \frac{\beta f_k^2}2\mathbf 1_{t\times t}
    \\
    \mathbf r_k =~& r_k\mathbf 1_{t\times 1}
\end{align}
RS parametrization of $\mathcal K_k(\mathbf Q_k)$ and $\mathcal P(\mathbf S_k, \mathbf r_k)$ dos not admit an analytic solution for $t>0$ on the real line.
By Taylor expansion around $t=0$, the terms can be sufficiently approximated up to order $t^2$.
Computing higher order terms is possible but tedious, especially for $\mathcal K_k(\mathbf Q_k)$.
The computations reveal that any integer power of the inverse temperature $\beta^i \forall i\in \mathbb Z^+$ is scaled by a corresponding $t^i$. 
Where $\mathbb Z^+$ is the set of positive integer numbers excluding $0$.
The RS parametrized exponents can therefore be sorted by their scaling in $\beta$ \eqref{eq:moments_function_isolated_spin_glass}.
\begin{align}
    f_{Z,k}^\mathrm{RS}(\beta,t)
    \propto
    \int& e^{-t \mathcal B_k^{(0)} - t\beta \mathcal B_k^{(1)} - t^2\beta^2 \mathcal B_k^{(2)} + \Omega(t^3)} \mathrm dP^\mathrm{RS}_k  
    \label{eq:moments_function_isolated_spin_glass}
\end{align}
The terms $\mathcal B_k^{(i)}$ are functions of the RS parameters $\chi_k$, $q_k$, $e_k$, $f_k$ and $r_k$ and $\mathrm dP_k^{RS}$ indicates integration over all said parameters.
In order to compute the first and second cumulant of the decision metrics, \eqref{eq:SCGF_mean_free_energy} and \eqref{eq:SCGF_var_free_energy} are evaluated at $\beta\uparrow\infty$.
Equation \eqref{eq:mean_isolated_distance_pre_limits} states the first cumulant after differentiating the SCGF $G(\beta,t)$ w.r.t. $t$.
\begin{align}
    \frac{\overline{D_{a;k}}}{M}
    =~& \lim_{\beta\uparrow\infty} \lim_{t\downarrow 0 }
    \frac{1}{M}
        \int 
            \frac{
                \frac{\mathcal B_k^{(0)}}{\beta} + \mathcal B_k^{(1)} + 2t\beta \mathcal B_k^{(2)} + \Omega(t^2)
    }{
        \int 
        e^{
            -t \mathcal B_k^{(0)} - t\beta \mathcal B_k^{(1)} - t^2\beta^2 \mathcal B_k^{(2)} + \Omega(t^3)
        }
       dP^\mathrm{RS}_k     
    }
    \nonumber
    \\
    &\times e^{
        -t \mathcal B_k^{(0)} - t\beta \mathcal B_k^{(1)} - t^2\beta^2 \mathcal B_k^{(2)} + \Omega(t^3)
    }
    \mathrm dP^\mathrm{RS}_k           
    \label{eq:mean_isolated_distance_pre_limits}
\end{align}
Typically, the integral in \eqref{eq:mean_isolated_distance_pre_limits} is solved in the limit $M\uparrow\infty$. In this case however not all terms scale with $M$ due to the finite-value constraint on the $L1$-Norm of the microstate: $\delta_{||\mathbf v||_1}$.
We merely enforce large \textit{but finite} $M,N$ such that the asymptotic equalities hold with sufficient accuracy.
Using a non-rigorous reordering of the limits, saddle-point integration is applied with $\beta$ as the large parameter. 
In order to facilitate the limit exchange, we omit all exponent terms in $\Omega(t^2)$ as they vanish fast enough for small $t$. Additionally we omit all scaling terms in $\Omega(t)$ as their contribution vanishes for $t\downarrow 0$ and there are no exponent terms in $\mathcal O\left(\frac{1}{t}\right)$.
The contribution of $\mathcal B_{k}^{(0)}$ grows insignificant at large $\beta$ and can hence be neglected in \eqref{eq:isolated_spin_glass_first_cumulant_limit_exchange} with vanishing error.
\begin{equation}
    \frac{1}{M}\overline{D_{a;k}}
    =
    \lim_{t\downarrow 0} \lim_{\beta\uparrow\infty}\frac{1}{M} \int \mathcal B_k^{(1)}\frac{\exp\{-t\beta\mathcal B_k^{(1)}\}}{\int \exp\{-t\beta\mathcal B_k^{(1)}\} \mathrm dP_K^\mathrm{RS}}
    \mathrm dP_K^\mathrm{RS}
    \label{eq:isolated_spin_glass_first_cumulant_limit_exchange}
\end{equation}
The fraction in  \eqref{eq:isolated_spin_glass_first_cumulant_limit_exchange} admits a \textit{Boltzmann-Gibbs distribution} with Hamiltonian $t\mathcal B_{k}^{(1)}$ and hence exhibits a ground-state property at $\beta\uparrow \infty$.
The integral is asymptotically solved in \eqref{eq:solved_mean_isolated_spin_glass}.
\begin{equation}
    \frac{1}{M}\overline{D_{a;k}}
    = \frac{1}{M}{\mathcal B_k^{(1)}}(\chi_k^*, q_k^*,e_k^*,f_k^*,r_k^*)
    \label{eq:solved_mean_isolated_spin_glass}
\end{equation}
Where $\chi_k^*$, $q_k^*$, $e_k^*$, $f_k^*$ and $r_k^*$ are the RS parameters at the extrema of $\mathcal B_k^{(1)}$, which is attained by setting all partial derivatives w.r.t. to the RS parameters to zero.
The variance \eqref{eq:solved_var_isolated_spin_glass} is obtained in a similar fashion, starting from the second order derivative of \eqref{eq:moments_function_isolated_spin_glass} w.r.t. $t$.
\begin{equation}
    \mathrm{Var\left(\frac{1}{M}D_{a;k}\right)}
    =
    -\frac{2}{M^2}\mathcal B_k^{(2)}(\chi_k^*, q_k^*,e_k^*,f_k^*,r_k^*)
    \label{eq:solved_var_isolated_spin_glass}
\end{equation}
The extrema RS parameters of \eqref{eq:solved_mean_isolated_spin_glass} and \eqref{eq:solved_var_isolated_spin_glass} are identical.

\subsection{The Coupled Spin Glass}
The sum variance $\mathrm{Var}(D_{a;a}+D_{a;b})$ is computed by \textit{coupling} the spin glasses: $\tau_a=\tau_b=1$.
The gaussian integrals remain virtually identical up to some minor changes in the determinant terms \eqref{eq:isolated_determinant_terms}.
It is only in the expectation over $\mathbf X_a$ and $\mathbf X_b$ \eqref{eq:joint_average_over_X} that the coupling is characterized by matrix $\mathbf C_{12}$, a function of $\tilde{\mathbf X}_a$ and $\tilde{\mathbf X}_b$.
\begin{align}
    \Xi_X
    =~&
    \int\int
    e^{
        -\alpha N\theta_{a,1} \mathbf u_{a,1}^T\mathbf C_{11} \mathbf u_{a,1}
        -2\alpha N\sqrt{\theta_{a,1}\theta_{b,1}} \mathbf u_{a,1}^T\mathbf C_{12} \mathbf u_{b,1}
    }
    \nonumber
    \\
    \times& e^{
        -2\alpha N\theta_{b,1} \mathbf u_{b,1}^T\mathbf C_{22} \mathbf u_{b,1}
    }
    \nonumber
    \\
    \times &e^{
        -\alpha N\sum_{k}\sum_{r=2}^t \theta_{k,2}\mathbf u_{k,r}^T\tilde{\mathbf X}_k\mathbf H_k\mathbf H_k \tilde{\mathbf X}_k\mathbf u_{k,r}
    }
    \mathrm d \mu(\mathbf U_a)\mathrm d \mu(\mathbf U_b)
    \label{eq:joint_average_over_X}
\end{align}
where $\mathbf C_{11}$ and $\mathbf C_{22}$ are functions of $\tilde{\mathbf X}_a$ and $\tilde{\mathbf X}_{b}$, respectively.
This spherical integral can be interpreted as a set of \textit{modified} (in the sense of \eqref{eq:modified_HCIZ_integral}) HCIZ integrals coupled through $\mathbf C_{12}$.
The derivations to solve \eqref{eq:joint_average_over_X} are omitted due to space constraints.
Instead we present our result as a proposition followed by two claims.
We have obtained a partial proof of the proposition, the claims are based on the results of \cite{guionnet_fourier_2005} and the corresponding modification used in the isolated spin glass computation.
\begin{proposition}[Rank-1 Operator Valued Free Fourier Transform]
    Consider a spherical integral of the form:
    \begin{align*}
    I_N(\{\theta_k\},\mathbf Z)
    =
        \int\int e^{-N\theta_a \mathbf u_{a,1}^T \mathbf C_{11}\mathbf u_{a,1} - N\theta_b\mathbf u_{b,1}^T\mathbf C_{22}\mathbf u_{b,1}}
        \\
        \times e^{-2N\sqrt{\theta_a\theta_b}\mathbf u_{a,1}^T\mathbf C_{12}\mathbf u_{b,1}}
        \mathrm d \mu(\mathbf U_a)\mathrm d \mu(\mathbf U_b)
    \end{align*}
    with $\mu(\cdot)$ being the Haar measure of the orthogonal group, $\theta_a, \theta_b\in \mathbb R^+$ and $||\mathbf C_{kl}||_\infty<\infty\forall k,l$.
    Furthermore, assume that the $2\times2$ block matrix $\mathbf Z$ is R-cyclic in the sense of \cite{speicher_free_2009}.
    \begin{equation*}
        \mathbf Z = \begin{pmatrix}
            \mathbf C_{11} & \mathbf C_{12} \\
            \mathbf C_{12}^T & \mathbf C_{22}
        \end{pmatrix}
    \end{equation*}
    Then, $I_N(\{\theta_k\},\mathbf Z)$ is asymptotically solved by:
    \begin{equation*}
        \lim_{N\uparrow\infty}
        I_N(\{\theta_k\},\mathbf Z)
        =\exp\left\{
            -\frac{N}{2} \int_0^1 \mathrm {tr}\left\{
                R_\mathbf Z^{\mathcal D_2}\left(-\mathbf P w\right)\mathbf P
            \right\}
            \mathrm d w
        \right\}
    \end{equation*}
    where $\mathbf P = \mathrm{diag}\{2\theta_a,2\theta_b\}$ and $R_\mathbf Z^{\mathcal D_2}(\cdot)$ is the operator valued R-transform of $\mathbf Z$ on the algebra of $2\times 2$ diagonal matrices: $\mathcal D_2$.
\end{proposition}
\begin{claim}[Extension to Rank-$\nu$]
    Proposition $1$ extends to the case of finite rank $\nu>1$ by the same argument as the Free Fourier Transform extends to rank $\nu\leq\rho \sqrt{N}$ for finite $\rho$.
    To reduce notational overhead, the mathematical definition is given in conjunction with Claim $2$.
\end{claim}
\begin{claim}[Modification of the Operator Valued Free Fourier Transform]
    The modification of the Free Fourier Transform holds also in the Operator Valued Case. 
    Consider a rank $\nu$ spherical integral of form:
    \begin{align*}
    &I_N(\{\theta_{k,r}\},\mathbf Z)
    =
        \int 
        \prod_{r=1}^\nu
        e^{-N\theta_{a,r} \mathbf u_{a,r}^T \mathbf C_{11,r}\mathbf u_{a,r}^T }
        \\
        &\times e^{- N\theta_{b,r}\mathbf u_{b,r}^T\mathbf C_{22,r}\mathbf u_{b,r}-2N\sqrt{\theta_{a,r}\theta_{b,r}}\mathbf u_{a,r}^T\mathbf C_{12,r}\mathbf u_{b,r}}
        \mathrm d \mu(\mathbf U_a)\mathrm d \mu(\mathbf U_b)
    \end{align*}
    Extending all assumptions of Proposition $1$ to accommodate $\nu>1$, the integral is asymptotically solved by:
    \begin{align*}
        &\lim_{N\uparrow\infty}
        I_N(\{\theta_{k,r}\},\mathbf Z)
        \nonumber
        \\&=\exp\left\{
            -\frac{N}{2} \sum_{r=1}^\nu\int_0^1 \mathrm {tr}\left\{
                R_{\mathbf Z_r}^{\mathcal D_2}\left(-\mathbf P_r w\right)\mathbf P_r
            \right\}
            \mathrm d w
        \right\}
    \end{align*}
\end{claim}
Based on Claim $1$, we are able to approximate $\Xi_X$ with sufficiently low error at large $M$ and $N$. 
Notably, the operator valued condition is only required for the $r=1$ terms as $\mathbf Z_r\forall r>1$ is diagonal with free on-diagonal entries \cite{speicher_free_2009}.
The trace of $R_{\mathbf Z_r}^{\mathcal D_2}$ collapses hence on the sum of scalar R-transforms for all $r>1$. 
\begin{align}
    \Xi_X \doteq
    \exp\left\{
        -\frac{\alpha N}{2} \int_0^1 \mathrm {tr}\left\{
            R_\mathbf Z^{\mathcal D_2}\left(-\alpha\mathbf P_1 w\right)\mathbf P_1
        \right\}
        \mathrm d w
    \right\}
    \nonumber
    \\
    \times\exp\{-\alpha N(t-1)\sum_{k\in\{a,b\}}\int_0^1 \theta_{k,2}R_{XJX}(-2\alpha w \theta_{k,2})\mathrm d w\}
\end{align}
$R_{XJX}(\cdot)$ is the R-transform of $\frac{1}{M}\tilde{\mathbf X}_k^T\mathbf H_k^T\mathbf H_k\tilde{\mathbf X}_k$. While a closed-form solution of $R_{XJX}(\cdot)$ exists, the operator valued $R_{\mathbf Z}^{\mathcal D_2}(\cdot)$ does not admit an analytic form. 
It is however possible to compute its Taylor expansion around $t$ up to $t^2$, which we once again omit to keep the notation compact. 
The sub-matrices of $\mathbf Z$ as well as the $\theta$ values are functions of the eigenvalues of $\mathbf V_a^T\mathbf V_a$ and $\mathbf V_b^T\mathbf V_b$. The same holds for the log-determinant terms originating from the gaussian integrals. 
Hence, the subshell definition in \eqref{eq:subshell_definition} remains unchanged and so do the subshell-density terms $\mathcal I(\mathbf Q_k)\forall k\in\{a,b\}$. 
The Taylor expanded terms are reordered to their scaling in $\beta$ as indicated in  $\eqref{eq:moments_function_coupled_spin_glass}$.
\begin{align}
    f_{Z}^\mathrm{RS}(\beta,t)
    \propto
    \int& e^{-t \mathcal B^{(0)} - t\beta \mathcal B^{(1)} - t^2\beta^2 \mathcal B^{(2)} + \Omega(t^3)}
    \prod_{k}\mathrm dP^\mathrm{RS}_k
    \label{eq:moments_function_coupled_spin_glass}
\end{align}
The computation of the mean and variance is structurally identical to $\eqref{eq:solved_mean_isolated_spin_glass}$ and $\eqref{eq:solved_var_isolated_spin_glass}$.
In fact we observe that $\mathcal B^{(1)} = \mathcal B^{(1)}_a+\mathcal B^{(1)}_b$, \textit{the coupled spin glass decouples in the first cumulant}. The same does \textit{not} apply to $\mathcal B^{(2)}$ which is to be expected since the mean \eqref{eq:mean_decision_metric} decouples and the variance \eqref{eq:variance_decision_metri} does not.
    
\section{Results}
\input{dh_nsed_alpha_1_and_10.tex}
\begin{figure}[t]
    \centering
    \includegraphics[width=1\columnwidth, height=.5\columnwidth]{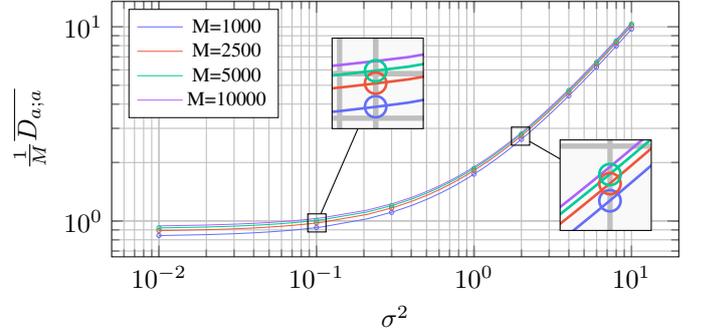}
    \caption{$\frac{1}{M}\overline{D_{a;a}}$ over $\sigma^2$ at $\alpha=1$}
    \label{fig:dh_nsed_mean}
    \squeezeup
\end{figure}
\input{ch_nsed_variance.tex}
\begin{figure}[t]
    \centering
    \includegraphics[width=1\columnwidth, height=.5\columnwidth]{ch_nsed_variance.tikz}
    \caption{$\mathrm{Var}\left(\frac{1}{M}D_{a;b}\right)$ over $\sigma^2$ at $\alpha=1$}
    \label{fig:ch_nsed_var}
    \squeezeup
\end{figure}
To verify that the replica computations produce correct results, the computed metrics are cross-validated to numerical values by Monte Carlo simulation. 
Throughout this comparison we plot the replica computation results as solid lines and the Monte Carlo simulations as markers. 
In Fig. \ref{fig:dh_nsed_mean} and Fig. \ref{fig:ch_nsed_var} the mean of the normalized squared distance to the DH and the variance of the normalized squared distance to the CH are plotted, respectively.
The numerical experiment is conducted for $M\in\{1000, 2500, 5000\}$ and $\alpha=\in\{1,10\}$. 
Equations \eqref{eq:solved_mean_isolated_spin_glass} and \eqref{eq:solved_var_isolated_spin_glass} predict the simulated values near perfectly. 
For $M=1000$ we carried out a total of $10^4$ Monte Carlo simulations whereas a minimum of $3500$ and $500$ simulations are conducted for $M=2500$ and $M=5000$, respectively due to the increased computational complexity.
As the room dimension $M$ increases the $L2$-Norm exhibits reduced ability to discriminate between near and far points \cite{goos_surprising_2001}. 
This explains why the mean squared distance in Fig. \ref{fig:dh_nsed_mean} converges to some value while the corresponding variance decreases as $M$ increases.
Furthermore, we note that the double logarithmic plots can be split into two areas.
For $\sigma^2>1$, both the mean and variance increase near linearly, with gradient one.
For $\sigma^2<1$, the mean and variance approach a constant which is equal to the values obtained in a hypothetical noise-free system. 
\input{accuracy_estimation_vs_numeric.tex}
\begin{figure}[t]
    \centering
    \includegraphics[width=1\columnwidth, height=.5\columnwidth]{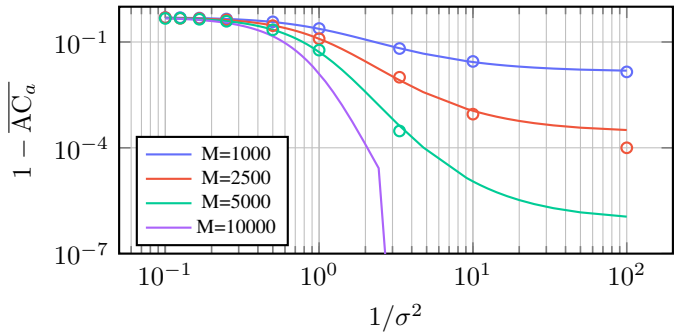}
    \caption{$1-\overline{\mathrm{AC}_a}$ over $\frac{1}{\sigma^2}$ at $\alpha=10$}
    \label{fig:accuracy_alpha_10}
    \squeezeup
\end{figure}
\begin{figure}[t]
    \centering
    \includegraphics[width=1\columnwidth, height=.5\columnwidth]{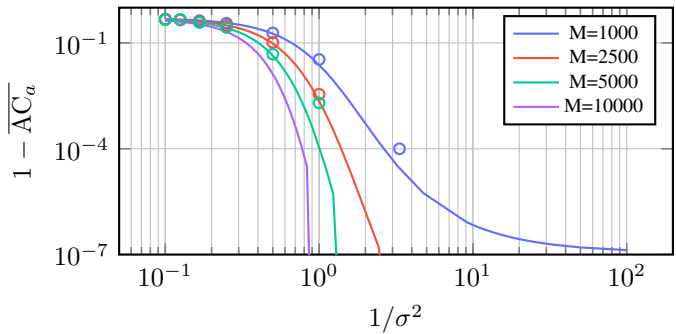}
    \caption{$1-\overline{\mathrm{AC}_a}$ over $\frac{1}{\sigma^2}$ at $\alpha=1$}
    \label{fig:accuracy_alpha_1}
    \squeezeup
\end{figure}
\begin{figure}[t]
    \centering
    \includegraphics[width=1\columnwidth, height=.5\columnwidth]{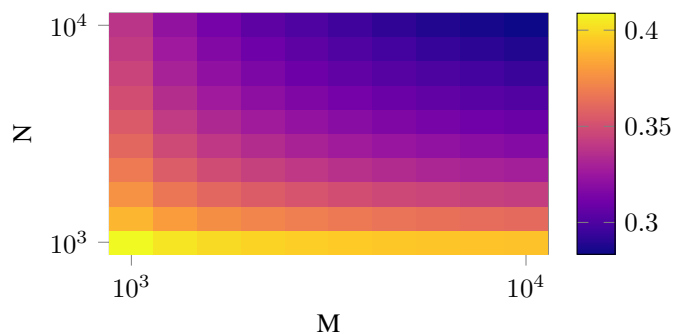}
    \caption{$1-\overline{\mathrm {AC}_a}$ over $M$ and $N$ at $\sigma^2=5$}
    \label{fig:mcr_heatmap}
    \squeezeup
\end{figure}
\\
The average accuracy \eqref{eq:accuracy_metric_average} is approximated by the partial integral over the moment-matched gaussian  \eqref{eq:accurcy_estimation}.
\begin{equation}
    \overline{\mathrm{AC}_a}
    \approx
    \int_{-\infty}^0\mathcal N(D_{a}; \overline{D_a}, \mathrm {Var}(D_a))\mathrm d D_a
    \label{eq:accurcy_estimation}
\end{equation}
Figure \ref{fig:accuracy_alpha_10} and \ref{fig:accuracy_alpha_1} depict the obtained estimate and simulated values at $\alpha=10$ and $\alpha=1$, respectively. 
The gaussian approximation is able to predict the phase transition form accuracy 0.5 ($1-\mathrm{AC}_a=0.5$), i.e. random classification, to near perfect classification in satisfactory detail.
In the asymptotic regime $\sigma^2\to 0$, the NCHC exhibits an error floor depending on the room dimension $M$ for constant $\alpha$.
The error floor is not visible for $M=10000$ at $\alpha=10$ and $M>1000$ for $\alpha=1$ due too limited numeric precision.
Furthermore, we observe a mismatch between the Monte Carlo simulation and replica computations at low misclassification rates.
This is due to the maximum number of simulations: $10^4$; limiting the error resolution to $10^{-4}$. 
For $M=5000$ and $\alpha=1$ the error accuracy limited to $2\cdot10^{-3}$.
As $\alpha$ increases, i.e. the length of the training sequence decreases, the critical SNR $1/\sigma^2_c$ at which the phase transition occurs increases and the error floor rises. 
This implies that low $\mathrm{SNR}$ may be compensated by an increased training sequence length and that the error floor is dependent on the number of training samples as it is induced by a sub-optimal sampling of the convex hull.
More notable however, higher noise power can be mitigated by increasing the number of antennas $M$.
A similar effect emerges in the heatmap of Fig. \ref{fig:mcr_heatmap}.
For constant $N$, increasing the room dimensions $M$ \textit{reduces} the error rate. 
This behavior might seem unintuitive at first, but can be explained twofold. 
\\
Firstly, consider that the overlap of the ellipsoids (Fig. \ref{fig:date_generation_illustration} (c)) is asymptotically governed by the singular values of $\mathbf H_a$ and $\mathbf H_b$ in the vicinity of $0$, as the two channel matrices are free and hence do not share a preferred direction. 
The relative overlap area therefore asymptotically tends to zero.
In other words: the probability of encountering a sample in the ambiguous area vanishes as $M\uparrow\infty$.
Apparently this effect is stronger than the finite sample size effect.
\\
Secondly, the number of users is constant while the number of antennas, i.e. the degrees of freedom (DoF) are increased, allowing for better classification performance.
This is not guaranteed to hold if the number of users were to grow in conjunction with the number of antennas. 

\section{Conclusion}
In this work, NCHC is applied to blind user identification in a massive MIMO system.
The BS is able to identify two UEs by comparing a received signal to $2N$ previously recorded transmissions of known origin without knowledge on their modulation format, coding, CSI or noise power.
The misclassification rate at finite, but large, number of antennas $M$ and training sequence length $N$ exhibits an error floor which decreases as $M$ \textit{or} $N$ are increased.
The NCHC is analyzed by the use of replica method for large but finite system size utilizing the inverse temperature $\beta$ as the large parameter.
We postulate the existence of an Operator Valued Free Fourier Transform based on an omitted partial proof.
The replica results are verified to numeric simulations which indicate that this proposition is correct. 
Future work includes but is not limited to: replica computations at the limit $M\uparrow\infty$ as $\alpha=\mathrm{const}$, providing a proof  of the Operator Valued Free Fourier Transform, scaling the number of users with the number of antennas and comparing the NCHC to the Bayes Optimal Classifier and the ANCHC.

\section*{Acknowledgment}
The authors would like to thank Roland Speicher for discussions on the operator valued R-transform, guidance on the corresponding computations and enabling a visit to the Department of Mathematics at Saarland University.

\bibliographystyle{IEEEtran}
\bibliography{references}

\end{document}

%% file: dh_nsed_alpha_1_and_10.tex
\pgfplotstableread{sigmaSquared DHnumeric1000alpha1 DHnumeric2500alpha1 DHnumeric5000alpha1 DHnumeric10000alpha1 DHnumeric1000alpha10 DHnumeric2500alpha10 DHnumeric5000alpha10 DHnumeric10000alpha10
0.01 0.8418932474445583 0.8922238176938443 0.9212448713873308 nan 0.9278008301681091 0.9434064896900942 0.9561185323061231 0.9676810199668044
0.1 0.924881726336103 0.9780393811844237 1.0069170670979908 nan 1.0147491177283272 1.031805522094549 1.0444442576352886 1.0555637023518458
0.3 1.1078460925735696 1.1660611007486794 1.1996504114374347 nan 1.210743355564015 1.2264268249134045 1.2399074209516652 1.252115451668497
1.0 1.743042712680526 1.8202749219485046 1.8643540200211735 nan 1.8851898448870485 1.9059138345947726 1.9216746912087739 1.9369470941877258
2.0 2.6356162968145194 2.7447028986731725 2.8047087924262937 nan 2.846256811818879 2.870222143152613 2.8913534069467093 2.912350681161247
4.0 4.415474071955541 4.588276690547727 4.696489199333728 nan 4.755391786115636 4.791705929764713 4.827674483166304 4.860775425287702
6.0 6.199243806312271 6.433521252425523 6.566303045723897 nan 6.660539833550096 6.7158428565973445 6.761219363677398 6.807682764643536
8.0 7.965795542742935 8.278683616200658 8.456646058250032 nan 8.563988401080948 8.631196879033217 8.6944187294815 8.755763982909992
10.0 9.742705998074515 10.125351409192008 10.328472363803694 nan 10.471836594577608 10.551896091306785 10.631594783687307 10.694630830226384
}\dataNumericNSEDdh
\pgfplotstableread{sigmaSquared DHreplica1000alpha1 DHreplica2500alpha1 DHreplica5000alpha1 DHreplica10000alpha1 DHreplica1000alpha10 DHreplica2500alpha10 DHreplica5000alpha10 DHreplica10000alpha10
0.01 0.8412571098725399 0.8920465150028429 0.9209712801597025 0.9432838072489996 0.9259613692072554 0.9433361013631544 0.9560766530331092 0.967319495607422
0.02 0.8505334981017126 0.901534941108847 0.9305820290989332 0.9529898656131094 0.9357632528586557 0.953131150958335 0.9658927916484848 0.9771637675616299
0.03 0.8598037510024837 0.9110193606243069 0.9401898689177439 0.9626938090831393 0.9455608453797576 0.9629232691693534 0.9757067532248573 0.9870064242761033
0.04 0.8690679942795055 0.9204998478421894 0.9497948540867296 0.9723956812104536 0.955354144849668 0.9727125164366737 0.9855185770420105 0.9968474968025575
0.05 0.8783263501330709 0.9299764796812908 0.9593970385038773 0.9820955180900048 0.9651433435581622 0.9824989510925194 0.9953283064631727 1.0066870164854338
0.06 0.887578940126813 0.9394493312346295 0.9689964751189828 0.9917933561264398 0.9749284925146708 0.9922826274587866 1.0051359839224345 1.0165250118411642
0.07 0.896825877879965 0.9489184743389871 0.9785932124876483 1.00148922865854 0.9847096650904718 1.0020636031245143 1.014941647929973 1.0263615119315073
0.08 0.9060672817639317 0.9583839766707377 0.9881873001062763 1.0111831740852428 0.9944869608437471 1.0118419262971732 1.0247453405950684 1.0361965429736422
0.09 0.915303417454192 0.96784590616345 0.9977787836754327 1.0208752234586933 1.0042604573286555 1.0216176643732653 1.034547091193638 1.0460301382816848
0.1 0.924534097044631 0.977304331509986 1.0073677143589939 1.030565411741689 1.0140302267036094 1.0313908541543133 1.0443469458547345 1.0558623188897618
0.11 0.9337595589921512 0.9867593168539625 1.0169541380428462 1.0402537714937306 1.0237963633613048 1.0411615500455256 1.0541449374325507 1.0656931127741827
0.21 1.0257485925798744 1.0811334772650891 1.112689829971261 1.1370435142095172 1.1212728144374866 1.138741512851937 1.1520297523648624 1.163930132016191
0.31 1.1173141415545205 1.1752254540403073 1.2082185229471254 1.233681942465295 1.2184562161791712 1.2361188961167024 1.2497622992145887 1.262053257102237
0.41 1.2085272485969956 1.2690805825089788 1.3035727933618557 1.330192179350006 1.3153974026155637 1.33332746901073 1.347367355736638 1.3600806512500192
0.51 1.2994443637161786 1.3627354145059236 1.3987788938667869 1.4265932102307215 1.4121363338452853 1.4303941517331293 1.44486480472936 1.4580269711515146
0.61 1.3901109689729234 1.4562196768583393 1.4938583522745246 1.5229006957616291 1.5087048733079982 1.527340571258348 1.5422707395143118 1.555904175133016
0.71 1.4805639585060324 1.5495577281619044 1.5888289506784767 1.6191275889085985 1.6051286420331508 1.6241843233570878 1.6395983203575346 1.6537219981048474
0.81 1.5708335279441559 1.6427697012363163 1.683705425396378 1.7152846621358198 1.7014284742612653 1.7209398257894641 1.7368583786139014 1.7514885861910356
0.91 1.6609445538407197 1.7358721774149124 1.7785000659856844 1.8113809323272625 1.7976214906267016 1.8176189703818317 1.8340598889864341 1.8492106802815695
1.1 1.8317901035195678 1.9125137342499872 1.9584200467369328 1.993822537903593 1.980142689615964 2.0011332036903218 2.018607370088071 2.0347795087096636
1.2 1.9215468200186476 2.005370311205413 2.053030145270895 2.0897813445412963 2.0760979201404064 2.0976414765856966 2.1156774259673456 2.1324013874380174
1.3 2.011208793345208 2.098160805233524 2.1475905233214614 2.185703036058135 2.1719897858953883 2.194104011078835 2.2127121889608405 2.2299963023968465
1.4 2.1007867443597092 2.190892577529029 2.2421066748453202 2.2815916825235734 2.2678255328395682 2.2905259556391773 2.3097156221291457 2.327567231352062
1.5 2.190289849691336 2.2835719646615265 2.3365833230323565 2.377450795491432 2.3636112888147633 2.3869117652889122 2.4066910944336275 2.425116792916259
1.6 2.279726019956312 2.3762044258740675 2.431024563328889 2.4732834152288365 2.4593526154547893 2.483265211717793 2.5036415346171976 2.52264716516627
1.7 2.3691020917977514 2.46879473478098 2.525433955657951 2.5690921901906556 2.5550538773609732 2.579589620226187 2.6005694858649195 2.6201603108260234
1.8 2.458424046329296 2.5613469438756185 2.6198146006521115 2.6648794369812943 2.650719138633227 2.6758878583796397 2.697477163522529 2.7176578967540936
1.9 2.5476970400333507 2.653864765838888 2.7141692238704556 2.7606471838533158 2.7463519492707507 2.772162392099526 2.79436650276765 2.815141412664368
2.0 2.636925675610074 2.746351351957312 2.808500209694732 2.85639721450253 2.8419552686669998 2.8684155808871368 2.8912391980930088 2.9126121555729076
2.1 2.726113956322437 2.8388095030539993 2.90280968417992 2.9521311085926385 2.9375318526655856 2.9646492095733663 2.988096750889324 3.0100712718414844
2.2 2.8152654324268944 2.9312417087210036 2.9970995168425016 3.0478502772350367 3.033084055445224 3.0608650501229353 3.084940476535879 3.107519776028939
2.3 2.904383257711654 3.023650175837143 3.0913713729243018 3.143555935478109 3.1286139822277654 3.15706462656202 3.181771589976874 3.2049585711128543
2.4 2.993470236434027 3.116036871072767 3.185626733954265 3.2392492228229424 3.2241235043182885 3.2532492888571896 3.278591114624557 3.3023884615166095
2.5 3.0825288731708618 3.208403548399376 3.2798669254744253 3.3349311251103857 3.319614288512384 3.349420272135295 3.37539998923509 3.399810166049753
2.6 3.171561409221024 3.3007517947229967 3.3740931393190836 3.4306025327219687 3.4150878248307888 3.445578641084423 3.4721990586095344 3.4972243279961313
2.7 3.2605698582546814 3.3930830154790903 3.468306372433939 3.5262642477010475 3.5105454533942804 3.541725352310046 3.568989078330961 3.594631524818533
2.8 3.3495560309840693 3.4853984867279646 3.5625078087727386 3.6219169924557617 3.605988379874265 3.637861316305413 3.6657707271558313 3.6920322864852446
2.9 3.4385215539885547 3.577699361743201 3.656698101049351 3.717561420141801 3.7014176917964177 3.7339873119212537 3.7625446180477384 3.789427077789519
3.0 3.5274679326508265 3.6699866837702175 3.750878112503173 3.8131981229065746 3.796834372607507 3.8301040561328374 3.8593113076213155 3.886816326164324
3.1 3.61639649563798 3.7622613259141304 3.8450485616134182 3.9088276390065237 3.8922393136359585 3.926212199292708 3.956071300944417 3.984200419588584
3.2 3.7053084608383906 3.8545242970081195 3.9392101003128266 4.004450458184895 3.987633324897781 4.022312332634597 4.052825057295617 4.081579703543981
3.3 3.794204970014834 3.9467763120563726 4.033363323546891 4.100067024316027 4.083017144015637 4.118404994742232 4.149572995173158 4.178954515157815
3.4 3.883086990400725 4.03901808752611 4.127508774955734 4.19567775157733 4.178391442597166 4.214490677304444 4.246315496740891 4.276325140574932
3.5 3.9719554728101696 4.1312502810201295 4.221646952863787 4.291283010488579 4.273756834975477 4.31056982941884 4.34305291147528 4.3736918481203
3.6 4.060811270578581 4.223473494181459 4.315778312270429 4.386883143714804 4.36911388487362 4.406642862949141 4.439785560191786 4.471054884224058
3.7 4.1496551671964825 4.315688279589471 4.4099032729897045 4.482478466255383 4.464463109700992 4.502710156179803 4.53651373698817 4.568414475306103
3.8 4.238487852645465 4.407895145747079 4.504022220931143 4.578069267817892 4.559804985119871 4.598772057370016 4.633237712583718 4.665770829629523
3.9 4.327310046889654 4.500094561062836 4.598135511344991 4.673655815577905 4.655139930201568 4.694828887102989 4.729957736380814 4.763124139079576
4.0 4.41612233328028 4.592286958075247 4.69224347188861 4.769238356157843 4.750468397782877 4.790880941100594 4.826674038530681 4.860474558680016
4.1 4.504925266231364 4.684472735532299 4.786346407059068 4.864817117345137 4.845790731225634 4.886928492724184 4.9233868317498315 4.957822303278114
4.2 4.593719393155131 4.776652263674539 4.880444597176872 4.960392310687588 4.941107274671459 4.982971794913795 5.020096490986351 5.055167503876044
4.3 4.682505186309471 4.868825886514763 4.974538302829909 5.05596413180397 5.036418348239182 5.0790110821490595 5.116802664595698 5.152510277347698
4.4 4.771283095483301 4.960993922557896 5.06862776590891 5.151532762586312 5.13172424895027 5.17504657203603 5.2135060561988675 5.24985076937176
4.5 4.860053539108467 5.053156668193117 5.162713211428612 5.247098367938182 5.227025252797026 5.271078466607188 5.310206645155015 5.347189101237026
4.6 4.9488169068162255 5.145314398541619 5.25679484889973 5.34266111329353 5.3223216164925 5.3671069543453305 5.406904578022141 5.444525385728278
4.7 5.037573561193185 5.237467372017918 5.350872873447658 5.438221141729647 5.417613579233866 5.46313221040408 5.503599991329736 5.541859728123688
4.8 5.126323841533048 5.3296158281478725 5.444947467006577 5.5337785894082065 5.512901363987699 5.559154398316881 5.600293012456228 5.639192226470703
4.9 5.215068062768111 5.42175999008835 5.539018800012131 5.629333585258388 5.6081851791482675 5.655173670798207 5.696983760390108 5.736522972412316
5.0 5.303806522852474 5.5139000687878905 5.63308703141348 5.7248862482428216 5.703465219344085 5.75119017063492 5.793672346341418 5.833852051724825
5.1 5.392539497526099 5.6060362599251174 5.727152310219374 5.8204366908545575 5.798741666771926 5.847204031448674 5.8903588744205955 5.931179544579642
5.2 5.481267246315824 5.698168747264824 5.8212147760346875 5.9159850178475555 5.894014692116484 5.943216377632235 5.987043442310962 6.028505526192373
5.3 5.569990012596178 5.790297703301851 5.9152745590504106 6.011531327708441 5.989284455461291 6.039225368752984 6.083726141467009 6.125830067054696
5.4 5.658708024945088 5.88242329002704 6.009331785788072 6.107075712951488 6.0845511070608165 6.135230993750317 6.180407057840657 6.223153233396886
5.5 5.747421498152496 5.974545659602342 6.103386569428313 6.202618260514032 6.179814723876028 6.231235475642861 6.277086272182557 6.320475087461922
5.6 5.836130634326866 6.066664954788692 6.1974390183885975 6.298159052248174 6.275075566612632 6.327237872398087 6.373763860337591 6.417795687794604
5.7 5.924835625381316 6.158781311677162 6.291489235085957 6.393698165167875 6.370333696917289 6.42323827576389 6.470439893793808 6.5151150895176775
5.8 6.013536651837686 6.250894856832835 6.385537316174975 6.489235671925115 6.465589232568527 6.519236772305074 6.567114439963301 6.612433344621836
5.9 6.102233880086445 6.343005710142984 6.4795833527020426 6.584771641022101 6.560842284638703 6.615233443558733 6.663787562464701 6.709750502121144
6.0 6.190927469856586 6.435113984859134 6.5736274304783 6.680306137105245 6.656092957952022 6.711228366520295 6.760459321296523 6.8070666082533515
6.1 6.279617571962708 6.5272197879230385 6.667669630997763 6.775839221174537 6.751341351495567 6.807221613923458 6.857129773219371 6.904381706659208
6.2 6.368304329138173 6.619323220533297 6.761710030902125 6.871370951335189 6.846587558959685 6.9032132545686915 6.953798971780247 7.001695838680115
6.3 6.45698787632247 6.711424378283197 6.855748703486761 6.966901382125644 6.941831668862741 6.9992033533889 7.050466967785488 7.099009043406918
6.4 6.54566834150889 6.803523351748384 6.949785717223508 7.062430565329701 7.037073765091877 7.095191948511752 7.147133809283329 7.196321357886358
6.5 6.634345845796523 6.895620227336268 7.04382113738668 7.157958550029837 7.132313927084096 7.191179146819614 7.24379954177155 7.2936328172064275
6.6 6.723020504130424 6.987715086221244 7.137855025860171 7.253485382787655 7.227552230110882 7.2871649788342 7.340464208414074 7.3909434460206365
6.7 6.811692425577016 7.079808005724427 7.231887441395402 7.349011107778337 7.322788745562572 7.383149497078893 7.4371278501714135 7.488253293376424
6.8 6.900361713174437 7.171899059273137 7.325918439767568 7.444535766949962 7.418023541314158 7.4791327514145625 7.533790505927843 7.585562380431199
6.9 6.989028465938978 7.263988316658004 7.419948073983664 7.540059400184041 7.513256681860441 7.575114789218031 7.630452212650915 7.682870735709121
7.0 7.0776927769338975 7.356075844126928 7.513976394424261 7.6355820453441545 7.6084882285107 7.67109565553037 7.72711300545728 7.780178386278315
7.1 7.166354722655806 7.448161704965824 7.608003449242955 7.731103738506827 7.7037182395030035 7.767075393208288 7.823772917819281 7.877485357910478
7.2 7.25501441471222 7.540245959170735 7.7020292837673585 7.826624514023148 7.7989467705321776 7.863054043039521 7.920431981509225 7.974791675132604
7.3 7.343671919624603 7.63232866402814 7.796053941636489 7.9221444045946905 7.894173874389921 7.959031643879109 8.017090226852225 8.072097361310323
7.4 7.432327315164547 7.724409874090003 7.890077464304345 8.017663441403855 7.989399601467247 8.055008232752257 8.11374768273022 8.16940243871259
7.5 7.520980674303451 7.816489641394494 7.984099891313102 8.113181654188828 8.084623999820286 8.150983844969335 8.210404376686617 8.26670692854992
7.6 7.609632067073619 7.908568014973037 8.078121260378284 8.208699071196737 8.1798471152914 8.246958514197557 8.30706033500112 8.364010851133925
7.7 7.698281560119031 8.000645043471671 8.17214160751372 8.304215719775005 8.275068991641628 8.342932272526037 8.40371558274823 8.461314225779411
7.8 7.786929217647904 8.092720771557842 8.266160967119554 8.399731625673915 8.37028967067001 8.438905150677874 8.500370143911242 8.55861707094907
7.9 7.875575100711434 8.184795242509553 8.360179372058132 8.495246813624556 8.465509192315581 8.534877177985829 8.597024041380015 8.655919404299484
8.0 7.964219266638917 8.27686849761494 8.454196853797747 8.590761307275836 8.560727594570913 8.630848382531017 8.693677297050286 8.753221242700137
8.1 8.052861772440513 8.368940576648647 8.548213442396321 8.686275129251332 8.655944914351158 8.726818791177594 8.790329931856203 8.850522602291532
8.2 8.14150267131191 8.461011517408974 8.642229166634728 8.7817883011919 8.75116118641845 8.82278842960835 8.88698196585069 8.94782349852529
8.3 8.230142015088475 8.553081356179192 8.736244054082668 8.877300843850211 8.846376444236027 8.918757322426297 8.983633418214264 9.045123946210081
8.4 8.3187798521899 8.64515012771917 8.830258131142147 8.97281277705939 8.94159071987256 9.014725493200475 9.080284307340582 9.142423959516767
8.5 8.407416229584715 8.737217865344373 8.92427142312116 9.06832411989849 9.036804044054174 9.110692964517192 9.176934650841268 9.239723552030954
8.6 8.49605119233751 8.829284600999998 9.018283954286558 9.163834890601139 9.132016446228358 9.206659758032185 9.273584465611119 9.337022736787068
8.7 8.584684783687793 8.921350365330595 9.112295747916923 9.259345106677735 9.227227954661958 9.30262589451763 9.370233767851616 9.434321526273338
8.8 8.67331704508161 9.013415187745514 9.20630678742908 9.354854784928579 9.322438596467785 9.398591393912712 9.466882573111912 9.53161993248826
8.9 8.761948016368573 9.105479096480412 9.300317171555717 9.450363941496082 9.417648397672613 9.494556275341427 9.563530896305258 9.628917966938829
9.0 8.850577735869313 9.197542118655162 9.394326882546844 9.54587259187189 9.512857383275016 9.59052055718802 9.660178751794874 9.726215640680534
9.1 8.939206240305468 9.28960428032833 9.488335940201004 9.641380750945496 9.608065577292082 9.686484257110124 9.756826153341237 9.823512964339463
9.2 9.027833565000947 9.3816656064367 9.582344363541988 9.736888433028785 9.703273002811034 9.78244739207963 9.853473114183489 9.920809948124418
9.3 9.116459743959867 9.473726121081443 9.676352170867093 9.832395651884108 9.798479682027946 9.878409978416144 9.950119647052137 10.018106601849972
9.4 9.205084809814755 9.56578584773779 9.770359379773186 9.927902420750636 9.893685636294059 9.974372031779621 10.046765764192184 10.115402934966449
9.5 9.293708794147578 9.657844808497678 9.864366007198507 10.023408752360638 9.988890886154632 10.070333567344974 10.143411477376556 10.21269895655816
9.6 9.382331727149463 9.749903024827221 9.958372069447549 10.118914658993454 10.084095451386592 10.166294599622212 10.24005679798168 10.30999467535395
9.7 9.470953638051224 9.841960517398764 10.052377582221752 10.214420152452272 10.179299351034086 10.262255142607762 10.3367017368929 10.407290099815341
9.8 9.559574554896859 9.93401730611281 10.146382560433373 10.309925244115549 10.274502603452202 10.358215209793189 10.43334630462169 10.504585238038187
9.9 9.648194504807806 10.026073410189566 10.240387019078854 10.405429944855921 10.369705226298247 10.454174814173287 10.529990511350812 10.601880097836755
10.0 9.736813513884622 10.118128848116095 10.334390972010326 10.500934265422933 10.464907236622482 10.550133968261964 10.626634366835276 10.69917468675908
}\dataReplicaNSEDdh

%% file: ch_nsed_variance.tex
\pgfplotstableread{sigmaSquared CHnumeric1000alpha1 CHnumeric2500alpha1 CHnumeric5000alpha1 CHnumeric10000alpha1 CHnumeric1000alpha10 CHnumeric2500alpha10 CHnumeric5000alpha10 CHnumeric10000alpha10
0.01 0.003543183433166508 0.0015169329564925782 0.0007290939691678711 nan 0.003761022296735028 0.0015637968172897976 0.000759602922932423 0.00039201038238401154
0.1 0.00382728891877282 0.0015764360335726035 0.0008476312719716272 nan 0.004198641112676205 0.0016992876827546244 0.0008517050431926521 0.000462959114317929
0.3 0.0047888183632813774 0.0019970165160243525 0.0009677115626152144 nan 0.005069904650294665 0.0020663504094313723 0.0010402034945518057 0.0005040805108358359
1.0 0.008879746897586749 0.0037151782968103397 0.00204666689020927 nan 0.009256113606158856 0.003863766634124577 0.0019120509011751885 0.0009864774586207936
2.0 0.017226961302399424 0.007504526545512924 0.0037470228721199206 nan 0.01888723546817772 0.007572156091850246 0.0038607434535204987 0.0019097155217409068
4.0 0.044729044026020404 0.018844247073271703 0.009684540467191027 nan 0.04965289998867206 0.019893059276473224 0.00991436617740149 0.0049269211331619545
6.0 0.08957581496087386 0.036384812274462774 0.020060296457700133 nan 0.09518507901218243 0.03885250116051964 0.01957188158608858 0.009670227491952232
8.0 0.14488086059630945 0.06183387341678781 0.032776951609278626 nan 0.15452110187867163 0.06441275415809855 0.03194768985710539 0.016113450056352008
10.0 0.21285828516501226 0.0888105074552783 0.0394635541185977 nan 0.23220171766047315 0.09354425983244141 0.046439722001125006 0.023106656923857827
}\dataNumericNSEDchVariance
\pgfplotstableread{sigmaSquared CHreplica1000alpha1 CHreplica2500alpha1 CHreplica5000alpha1 CHreplica10000alpha1 CHreplica1000alpha10 CHreplica2500alpha10 CHreplica5000alpha10 CHreplica10000alpha10
0.01 0.0035323763940099937 0.0014741732592774774 0.0007545333685576779 0.0003839642264892946 0.0037899470993816606 0.001535639806534271 0.0007753889298734278 0.0003911007361178834
0.02 0.003568018603198476 0.0014890267130168863 0.0007621304964277093 0.0003878279811449301 0.003828800454862831 0.0015512286455368943 0.0007832292139934894 0.0003950458709810498
0.03 0.003604013914604629 0.0015040268880766189 0.0007698025670180928 0.0003917298262864761 0.0038680380969165515 0.0015669713507214674 0.0007911468157884554 0.0003990298950158188
0.04 0.003640362770136062 0.0015191737781186524 0.00077754957752686 0.000395669761346482 0.003907659981280911 0.0015828679139655154 0.0007991417331101188 0.0004030528076264041
0.05 0.0036770648620925786 0.0015344673696590311 0.0007853715256764252 0.0003996477856689564 0.00394766606590772 0.0015989183274607072 0.000807213963964506 0.00040711460824108194
0.06 0.0037141201351743214 0.0015499076590612915 0.000793268409071108 0.0004036638986854818 0.003988056310900622 0.0016151225837936999 0.0008153635064391511 0.00041121529632019694
0.07 0.003751528530877066 0.0015654946291844113 0.0008012402256572257 0.000407718099798777 0.004028830678275336 0.0016314806752171249 0.0008235903587033236 0.00041535487135828814
0.08 0.0037892899974289822 0.0015812282754186248 0.000809286973362915 0.00041181038845231657 0.004069989131814498 0.0016479925963793917 0.0008318945189992142 0.0004195333328635831
0.09 0.003827404481565633 0.001597108589533632 0.0008174086495789666 0.0004159407641984876 0.004111531637244425 0.0016646583395297171 0.0008402759856682441 0.00042375068033842115
0.1 0.0038658719298793956 0.0016131355632132886 0.0008256052526996172 0.000420109226777087 0.004153458161368117 0.0016814778998389724 0.0008487347571433792 0.00042800691335051694
0.11 0.0039046922834765512 0.0016293091890130903 0.0008338767806970996 0.0004243157751158243 0.004195768672916294 0.0016984512707596089 0.0008572708318690291 0.0004323020314748812
0.21 0.004312260938288704 0.001799109838353873 0.0009207125768957255 0.00046847589924846886 0.004639987139232178 0.001876643450396985 0.0009468829697583782 0.0004773918117639997
0.31 0.00475501826920205 0.0019835692340714702 0.0010150392919918143 0.000516444223351775 0.005122578494117309 0.002070211852088789 0.0010442241742124477 0.0005263697461578107
0.41 0.005233009300541724 0.0021826833359610254 0.0011168559457375514 0.0005682204972038456 0.0056435271384061075 0.0022791535330080777 0.001149293678096292 0.0005792356179238543
0.51 0.005746217754071089 0.002396449376864693 0.001226161851672355 0.0006238045488497253 0.006202822522673752 0.0025034665003300775 0.0012620909606583812 0.0006359892786615737
0.61 0.006294631391688165 0.0026248653963683965 0.001342956537269812 0.0006831962568463015 0.006800457316319509 0.0027431493640728394 0.0013826156577548287 0.0006966306267496277
0.71 0.0068782412230097975 0.0028679299852920237 0.001467239657314028 0.0007463955329299773 0.007436426295358298 0.002998201131112237 0.0015108675092529392 0.0007611595844321737
0.81 0.007497040371553229 0.0031256421152599943 0.0015990109596522378 0.0008134023146254906 0.00811072565137459 0.0032686210766522516 0.0016468463238340188 0.0008295760993773911
0.91 0.008151024106703136 0.003398001014166822 0.0017382702574053898 0.0008842165514522373 0.008823352553695036 0.0035544086614297775 0.0017905519627607314 0.0009018801310710887
1.1 0.009490514143669507 0.003955832790330097 0.002023491864759509 0.0010292530200580903 0.010282928799143956 0.00413974153376049 0.002084879707337386 0.0010499677969247937
1.2 0.010246515078471715 0.0042706644902856345 0.0021844656515678856 0.00111110870286033 0.011106695159046753 0.004470093283510436 0.002250992608223765 0.0011335454787440317
1.3 0.011037690229594086 0.0046001413108529106 0.002352927034411702 0.0011967717400215014 0.011968783088922918 0.004815811532539193 0.002424832032209749 0.0012210105919654613
1.4 0.011864038189132413 0.00494426304467986 0.002528875963672657 0.0012862421180734917 0.012869191770155886 0.005176896124523252 0.002606397937930222 0.00131236312497735
1.5 0.012725557797953209 0.005303029514362667 0.0027123123862357818 0.0013795198251698381 0.013807920543386686 0.005553346934110849 0.002795690292544501 0.0014076030682768903
1.6 0.013622248145805125 0.00567644057386575 0.0029032362703557647 0.0014766048524531951 0.01478496887491781 0.005945163859195261 0.0029927090691484453 0.0015067304142135222
1.7 0.01455410848388095 0.006064496105340682 0.003101647586515564 0.0015774971926494955 0.015800336327392757 0.006352346815972475 0.0031974542459524474 0.0016097451565597698
1.8 0.01552113819804773 0.0064671960112845876 0.003307546311578052 0.0016821968397501411 0.01685402254082007 0.006774895736387543 0.0034099258048401228 0.001716647290158167
1.9 0.01652333677906279 0.006884540209990909 0.0035209324252233032 0.0017907037896320364 0.01794602721587897 0.007212810564908355 0.0036301237308555556 0.0018274368107604703
2.0 0.017560703800675596 0.007316528638325246 0.003741805911779958 0.001903018035324746 0.019076350102323142 0.007666091250330845 0.0038580480118818993 0.001942113714722054
2.1 0.018633238910726886 0.007763161230466835 0.0039701667578052715 0.0020191395759719616 0.020244990989659196 0.008134737754778633 0.0040936986367233405 0.002060677999075104
2.2 0.019740942066598363 0.008224437944852699 0.004206014949582926 0.002139068406592519 0.02145194970769429 0.008618750043935754 0.00433707559665574 0.002183129661319658
2.3 0.020883812502506784 0.00870035874392417 0.004449350478249956 0.002262804525709503 0.022697226089391145 0.009118128088819049 0.004588178884349748 0.0023094686991951396
2.4 0.022061850254991407 0.009190923588096988 0.004700173335697668 0.002390347930546784 0.02398082001276939 0.009632871864984436 0.004847008492978956 0.002439695110889682
2.5 0.02327505513804742 0.00969613245060493 0.004958483514228534 0.002521698619306198 0.025302731366200006 0.010162981351330334 0.00511356441717034 0.0025738088948517425
2.6 0.024523426988995418 0.010215985304907884 0.005224281008413385 0.0026568565908016346 0.02666296005419189 0.010708456529468436 0.005387846653026167 0.0027118100525028816
2.7 0.02580696567386548 0.010750482130181939 0.005497565815737403 0.0027958218426595745 0.02806150599209029 0.011269297383827847 0.005669855194724293 0.002853698577153722
2.8 0.027125671077711213 0.011299622908265763 0.00577833792437695 0.0029385943747329804 0.029498369107830213 0.011845503900424421 0.005959590039415176 0.0029994744704570407
2.9 0.028479543087429258 0.011863407622664648 0.006066597333603435 0.003085174185239587 0.03097354933822862 0.012437076067529123 0.006257051184241363 0.0031491377316132706
3.0 0.029868581646382952 0.012441836257903507 0.006362344042761366 0.0032355612735615437 0.03248704661968862 0.01304401387407491 0.006562238625865881 0.003302688359871684
3.1 0.03129278661981538 0.013034908812738704 0.006665578043505959 0.003389755638862952 0.03403886092981782 0.013666317311131009 0.0068751523621470014 0.0034601263541321685
3.2 0.032752157963378 0.013642625254105294 0.006976299335932648 0.003547757280733533 0.03562899219603472 0.014303986370500779 0.007195792390891282 0.003621451714118878
3.3 0.03424669550942019 0.014264985588210086 0.007294507917066924 0.003709566197915467 0.037257440390494966 0.014957021044884684 0.007524158710205462 0.0037866644391723334
3.4 0.03577639942148806 0.014901989796288186 0.007620203785206761 0.003875182390054333 0.038924205480169574 0.015625421327882104 0.007860251318486723 0.003955764528876818
3.5 0.03734126953519641 0.015553637875712074 0.007953386938269877 0.00404460585686282 0.04062928743331334 0.016309187213930938 0.008204070214180828 0.004128751982731957
3.6 0.03894130581615016 0.01621992981950349 0.00829405737452787 0.004217836597460751 0.04237268622465359 0.01700831869782239 0.008555615395752737 0.0043056268004166365
3.7 0.040576508222356415 0.016900865625196576 0.008642215093072938 0.004394874612044591 0.044154401829140155 0.017722815775014892 0.008914886862366436 0.00448638898159667
3.8 0.04224687672031334 0.017596445280819176 0.0089978600914863 0.004575719900334337 0.04597443422513981 0.018452678441267682 0.009281884612904877 0.004671038525964325
3.9 0.04395241128340169 0.018306668796694035 0.009360992369366725 0.004760372460793833 0.04783278339326996 0.01919790669319156 0.009656608646317127 0.004859575433165704
4.0 0.04569311188665185 0.019031536145536643 0.009731611924868638 0.004948832293551653 0.0497294493159939 0.01995850052732343 0.010039058961694726 0.005051999703129224
4.1 0.047468978852277084 0.019771047334678524 0.010109718757467447 0.005141099399290295 0.051664431977550815 0.020734459941515068 0.010429235560611977 0.005248311335352896
4.2 0.04928001146528862 0.020525202358502356 0.010495312865994722 0.005337173776898881 0.053637731363381194 0.02152578493130781 0.010827138437799505 0.005448510329816979
4.3 0.05112621004369596 0.021294001215908104 0.010888394250944473 0.005537055426311227 0.05564934746050156 0.02233247549520773 0.011232767594901303 0.005652596686285864
4.4 0.0530075745751739 0.022077443911254196 0.011288962909397168 0.005740744347093465 0.05769928025683599 0.023154531631141843 0.011646123031259338 0.0058605704046408484
4.5 0.05492410504101472 0.02287553042585646 0.011697018842740953 0.005948240539208602 0.05978752974214243 0.023991953336891216 0.012067204746344254 0.006072431484659872
4.6 0.056875801427759845 0.0236882607678816 0.012112562048483768 0.006159544002645136 0.061914095906116136 0.024844740610756876 0.0124960127400041 0.006288179926229587
4.7 0.058862663717237675 0.02451563493185177 0.012535592526800738 0.006374654740375743 0.06407897874026494 0.025712893450806067 0.012932547011192355 0.006507815729287154
4.8 0.0608846919092709 0.02535765293106933 0.012966110277716467 0.006593572745806986 0.06628217823626141 0.026596411855690578 0.013376807559829795 0.006731338893714971
4.9 0.06294188598463848 0.02621431474398558 0.013404115300090226 0.006816298022838796 0.06852369438641763 0.02749529582388113 0.013828794385467058 0.006958749419283179
5.0 0.06503424593017287 0.0270856203662599 0.013849607594436621 0.00704283057006984 0.07080352718388248 0.028409545354215377 0.014288507487874599 0.007190047305920454
5.1 0.06716177173925895 0.027971569803884534 0.014302587159532082 0.007273170388753341 0.0731216766222397 0.02933916044552246 0.014755946866661495 0.007425232553595858
5.2 0.06932446340426707 0.028872163056655187 0.014763053993710699 0.007507317477164901 0.07547814269579786 0.03028414109646055 0.015231112521856086 0.007664305162148327
5.3 0.07152232091742175 0.02978740012168753 0.015231008098867226 0.007745271834770773 0.07787292539911292 0.03124448731036498 0.015714004452658876 0.00790726513158521
5.4 0.0737553442710928 0.030717280999722145 0.015706449473891625 0.00798703346366157 0.08030602472755649 0.03222019907809404 0.016204622659193837 0.008154112461915403
5.5 0.0760235334624861 0.031661805690021694 0.01618937812001982 0.008232602361946116 0.08277744067557255 0.033211276402888554 0.01670296714089448 0.008404847152832382
5.6 0.07832688847657374 0.03262097418920193 0.016679794033386895 0.008481978529642737 0.08528717324008911 0.03421771928388365 0.017209037897831413 0.008659469204401618
5.7 0.08066540931754522 0.03359478649699894 0.017177697216566763 0.008735161968759484 0.08783522241643775 0.035239527720331086 0.017722834929742054 0.00891797861666089
5.8 0.08303909597237306 0.03458324261379618 0.017683087671556987 0.008992152676977892 0.0904215882016351 0.03627670171159285 0.018244358236593972 0.009180375389307447
5.9 0.0854479484441936 0.03558634253757761 0.018195965392667997 0.009252950655004782 0.09304627059087832 0.037329241257033234 0.01877360781785298 0.009446659522634508
6.0 0.08789196672051362 0.0366040862679198 0.018716330392573477 0.009517555900816333 0.09570926960826387 0.03839714635595305 0.019310583673726237 0.009716831016282903
6.1 0.09037115079995856 0.03763647380407132 0.019244182649436597 0.0097859684181758 0.0984105852001141 0.03948041700798404 0.01985528580392238 0.009990889870468953
6.2 0.09288550067971044 0.03868350514694212 0.019779522175095255 0.010058188205172429 0.10115021738771585 0.04057905321285513 0.020407714208429725 0.01026883609005204
6.3 0.095435016356587 0.03974518030783752 0.020322348967334113 0.010334215260760346 0.10392816616870333 0.04169305496951189 0.020967868886719275 0.010550669664912206
6.4 0.09801969782628843 0.04082149925481312 0.02087266302716252 0.010614049586003125 0.10674443154246079 0.04282242227777795 0.02153574983910812 0.010836390600236038
6.5 0.10063954508676436 0.041912462009498774 0.02143046435571349 0.010897691179524882 0.10959901350442344 0.043967155137165766 0.022111357065550133 0.011125998895570461
6.6 0.10329455814607694 0.04301806856772561 0.0219957529504404 0.01118514004310056 0.11249191205266335 0.04512725354819484 0.02269469056536991 0.011419494551254725
6.7 0.10598473697954455 0.044138318929274716 0.02256852881343856 0.01147639617662042 0.11542312718645187 0.046302717508766314 0.02328575033901458 0.011716877567334734
6.8 0.10871008159689152 0.04527321309637477 0.023148791941003 0.011771459579075522 0.1183926589043297 0.04749354701965751 0.023884536386848544 0.012018147943417315
6.9 0.11147059199160042 0.046422751062241634 0.023736542342615772 0.012070330249413813 0.12140050720375152 0.04869974208028205 0.024491048707634452 0.012323305679948928
7.0 0.11426626822658167 0.04758693283054126 0.02433178000650982 0.012373008188324276 0.12444667208268943 0.04992130269038651 0.025105287301553113 0.01263235077622527
7.1 0.11709711017860626 0.04876575839922922 0.024934504939044904 0.01267949339787486 0.12753115354026334 0.05115822884980938 0.025727252169014597 0.012945283232762972
7.2 0.11996311791629098 0.04995922776849365 0.025544717134969547 0.012989785874991035 0.1306539515749341 0.052410520558176174 0.02635694330962311 0.013262103049165532
7.3 0.12286429141636523 0.05116734096018554 0.026162416598944138 0.013303885621355425 0.1338150661848732 0.05367817781523339 0.026994360723522282 0.013582810225684078
7.4 0.12580063068574823 0.052390097935027274 0.026787603327849987 0.013621792636756478 0.13701449737015878 0.054961200620825365 0.027639504410468946 0.013907404762250926
7.5 0.12877213572955382 0.05362749870648549 0.02742027732616114 0.013943506922823442 0.14025224512838705 0.05625958897445349 0.028292374370368877 0.01423588665910648
7.6 0.13177880653751983 0.05487954328320154 0.028060438588183252 0.014269028474823791 0.1435283094590066 0.05757334287629226 0.02895297060336643 0.014568255915467081
7.7 0.13482064311635916 0.05614623165408843 0.02870808711601182 0.014598357295793554 0.14684269036130929 0.05890246232616726 0.029621293109309582 0.014904512532104284
7.8 0.1378976454668111 0.05742756382590908 0.029363222910403972 0.014931493386894616 0.15019538783228337 0.06024694732362853 0.030297341888090162 0.015244656508624407
7.9 0.14100981357280526 0.05872353979581829 0.03002584597206419 0.01526843674744263 0.1535864018733183 0.061606797868499756 0.030981116939815045 0.015588687845194776
8.0 0.14415714744081928 0.060034159564209114 0.03069595629879123 0.015609187373820988 0.15701573248171385 0.06298201396062361 0.03167261826443622 0.015936606541530358
8.1 0.1473396470722319 0.06135942313028371 0.03137355389186641 0.015953745270672007 0.16048337965904247 0.0643725955999596 0.03237184586174088 0.016288412597785386
8.2 0.15055731246889534 0.06269933049754682 0.03205863875032639 0.016302110433174685 0.16398934340327478 0.0657785427869321 0.033078799731420146 0.01664410601404661
8.3 0.15381014362243728 0.06405388166102388 0.03275121087419015 0.01665428286992216 0.16753362371308894 0.06719985552002607 0.03379347987429786 0.01700368679047802
8.4 0.1570981405327106 0.06542307663159634 0.03345127026619263 0.017010262570657377 0.17111622058849035 0.06863653380008831 0.034515886289408895 0.01736715492642093
8.5 0.16042130320503978 0.06680691539290566 0.03415881691934695 0.01737004954041081 0.17473713402892713 0.07008857762695017 0.035246018977781575 0.017734510422132303
8.6 0.16377963163219184 0.0682053979515292 0.03487385084001297 0.017733643778375874 0.17839636403299713 0.07155598700016112 0.03598387793819995 0.018105753278073476
8.7 0.16717312582055596 0.06961852430792953 0.03559637202933194 0.018101045282675463 0.18209391060019145 0.0730387619196318 0.03672946317138499 0.018480883493642228
8.8 0.17060178575878848 0.07104629446194423 0.03632638047950145 0.018472254057737912 0.18582977373042706 0.07453690238571319 0.0374827746772894 0.01885990106899481
8.9 0.17406561145420063 0.0724887084156312 0.03706387619805952 0.018847270101260232 0.18960395342298658 0.07605040839786675 0.03824381245507753 0.019242806004122738
9.0 0.17756460290736137 0.07394576616511972 0.0378088591807323 0.01922609341145946 0.19341644967718843 0.07757927995600854 0.03901257650575588 0.01962959829920869
9.1 0.18109876011299833 0.07541746772095896 0.038561329429074184 0.01960872399068214 0.19726726249287171 0.07912351706018834 0.039789066833186275 0.02002027795395097
9.2 0.18466808369065957 0.07690381306207793 0.03932128694104086 0.019995161836992483 0.20115639186913356 0.0806831197101352 0.04057328342890996 0.020414844968804007
9.3 0.18827257241536294 0.07840480221555636 0.04008873172182111 0.02038540695259745 0.20508383780590167 0.08225808790626749 0.04136522629661608 0.020813299342959227
9.4 0.19191222689322207 0.07992043515355629 0.040863663762659784 0.020779459334098315 0.20904960030267483 0.08384842164849762 0.04216489543711676 0.021215641077163123
9.5 0.19558704712306044 0.08145071188755008 0.04164608307307079 0.02117731898713794 0.21305367935904249 0.08545412093571193 0.042972290849578286 0.021621870171049815
9.6 0.19929703310245436 0.08299563242062197 0.042435989644249966 0.02157898590749925 0.2170960749740632 0.0870751857689075 0.04378741253470887 0.022031986624776102
9.7 0.20304218483901507 0.08455519675172536 0.04323338348512851 0.021984460093985612 0.22117678714838326 0.08871161614742919 0.044610260491957286 0.022445990438202286
9.8 0.20682250232106608 0.08612940487293555 0.044038264589469155 0.0223937415464907 0.22529581588122105 0.09036341207133684 0.04544083472144938 0.022863881611426586
9.9 0.21063798555666768 0.0877182568241583 0.044850632957184784 0.022806830270708848 0.22945316117192918 0.09203057354109698 0.046279135223138736 0.02328566014437846
10.0 0.21448863454208258 0.08932175254748677 0.04567048859555181 0.023223726259831018 0.23364882302056797 0.09371310055571036 0.04712516199729248 0.023711326037030057
}\dataReplicaNSEDchVariance

%% file: accuracy_estimation_vs_numeric.tex
\pgfplotstableread{1OverSigmaSquared M1000alpha1 M2500alpha1 M5000alpha1 M10000alpha1 M1000alpha10 M2500alpha10 M5000alpha10 M10000alpha10
100.0 0.0 0.0 0.0 nan 0.0142 0.0001 0.0 0.0
10.0 0.0 0.0 0.0 nan 0.0281 0.0009 0.0 0.0
3.3333333333333335 0.0001 0.0 0.0 nan 0.0652 0.01 0.0003 0.0
1.0 0.034 0.0035 0.002 nan 0.2407 0.1235 0.0588 0.014102564102564103
0.5 0.1911 0.1039622641509434 0.048 nan 0.3723 0.2898 0.2213 0.1428205128205128
0.25 0.362 0.3134090909090909 0.276 nan 0.4479 0.4238 0.3869 0.3541025641025641
0.16666666666666666 0.426 0.4015384615384615 0.378 nan 0.4761 0.4618 0.4406 0.4305263157894737
0.125 0.4551 0.4469444444444444 0.468 nan 0.4892 0.4677 0.47424242424242424 0.4369230769230769
0.1 0.4746 0.46714285714285714 0.454 nan 0.4913 0.4897 0.46525252525252525 0.47
}\dataAccuracyOverSigmaNumeric
\pgfplotstableread{1OverSigmaSquared M1000alpha1 M2500alpha1 M5000alpha1 M10000alpha1 M1000alpha10 M2500alpha10 M5000alpha10 M10000alpha10
100.0 1.3372302849660492e-07 9.10432742800114e-18 2.9381636589721425e-33 1.1129352046062265e-64 0.015378636937600155 0.0003191866073277538 1.1170879119206714e-06 3.426351188728547e-21
50.0 1.5956903242296498e-07 1.456563503782874e-17 7.55423147957133e-33 7.529147435235825e-64 0.01647094837957438 0.0003720738371158545 1.4786578150468169e-06 8.403328963315337e-21
33.333333333333336 1.90796534583833e-07 2.342883468310927e-17 1.9540270628301436e-32 5.124252775814652e-63 0.017613846762872032 0.00043215430774327365 1.944726941961836e-06 2.0278467262623202e-20
25.0 2.2856214386846311e-07 3.7869036195418536e-17 5.080354716039184e-32 3.5027232851178977e-62 0.018807422086018415 0.0005002279573209738 2.5415851633346026e-06 4.8153436700671014e-20
20.0 2.742695149703812e-07 6.147542891280021e-17 1.326483792498675e-31 2.4007327217128203e-61 0.020052201329339842 0.0005770867669843171 3.3010610191245314e-06 1.125330074299107e-19
16.666666666666668 3.296220345159088e-07 1.0017844321060424e-16 3.4751246758031477e-31 1.647176957558808e-60 0.021348214082830493 0.0006635642189378021 4.261414014974571e-06 2.588483542392729e-19
14.285714285714285 3.966823912719999e-07 1.637845249807423e-16 9.126727559462101e-31 1.1294905466663238e-59 0.022695440908187158 0.0007605355044631959 5.468305521247099e-06 5.861188025316828e-19
12.5 4.779479335356258e-07 2.6851421956663224e-16 2.400825088208059e-30 7.728915962582608e-59 0.02409384607220472 0.0008689112686497322 6.975907933751675e-06 1.3066575695692824e-18
11.11111111111111 5.717263882182463e-07 4.4119459658607904e-16 6.320243313046655e-30 5.269697411984378e-58 0.025543223728806783 0.0009896472454484315 8.847944486458473e-06 2.8685149553732436e-18
10.0 6.902084531791992e-07 7.2617251257719e-16 1.6637238520260068e-29 3.5748762051682366e-57 0.027043248551331948 0.001123715980433061 1.1159150658607547e-05 6.202272817479e-18
9.090909090909092 8.338728374166818e-07 1.1967328253241253e-15 4.3757410235401965e-29 2.409576323966789e-56 0.02859358604523583 0.0012721307934360554 1.3996345653411173e-05 1.3209423687951462e-17
4.761904761904762 5.5207612021408965e-06 1.7122318373403208e-13 5.635253027057076e-25 2.514105736691536e-48 0.04670480493756639 0.003780208549654104 0.0001019423427097165 1.1555090422917003e-14
3.2258064516129035 3.2697861644869125e-05 1.7345941621611622e-11 3.3849942745317516e-21 4.5520782401280675e-41 0.06881180846939142 0.008882248195584044 0.00048103505190215997 2.821493828109432e-12
2.4390243902439024 0.0001564003446237286 1.6344087671457427e-07 6.7858334568509674e-18 8.975411611294378e-35 0.09355105616201694 0.017376930683108124 0.0016182565102527192 2.6436668532378113e-05
1.9607843137254901 0.0005864819280248468 2.074297612519961e-06 4.2634510020151295e-15 1.941965590816238e-29 0.11953613329617355 0.029521195223757783 0.0041962445904519426 0.00014594115225274537
1.639344262295082 0.00174311876160507 1.6539708581704422e-05 9.109452835558461e-13 5.5735809105242e-25 0.14562291068721187 0.045020409865446955 0.008915238498531857 0.000559993321690869
1.4084507042253522 0.004229826698822977 8.874788707071683e-05 7.576336739737893e-11 2.767198939841251e-21 0.17097984837717023 0.0631950153799623 0.01626045656856511 0.0016268879907388838
1.2345679012345678 0.008674589230613736 0.0003441399393114012 5.3261934651568835e-06 3.1376493629951565e-18 0.1950739565576054 0.08320901545561916 0.026372919997333887 0.003818165777706315
1.0989010989010988 0.015530902925251468 0.001028235139961292 3.125790283969172e-05 1.050442840050672e-15 0.21760978845155868 0.10424583145295163 0.03906636416681593 0.007601487308874569
0.9090909090909091 0.03552522468257993 0.004820553306387115 0.00037757774970850213 4.9202238944055675e-12 0.2557764936733978 0.14466185944300938 0.06869651045597557 0.020239917883760224
0.8333333333333334 0.049261219728648734 0.008836882920955441 0.0009997364501757738 3.167619489424983e-05 0.27345270998662224 0.1652761760910961 0.08617204167724626 0.029862686714466326
0.7692307692307692 0.06464000832984451 0.014585300281779368 0.0022307250414127016 0.00011518825500266098 0.2895755848310225 0.18504624887563487 0.10427199834750797 0.04132081521249424
0.7142857142857143 0.08116083291846736 0.022140256420529955 0.00434308820384057 0.000335510270358489 0.30425828054534065 0.20380720383184872 0.12256763333471654 0.054308104139825106
0.6666666666666666 0.09835407220687198 0.03142648209769788 0.0075807679431956695 0.0008181229263314415 0.31762252145549474 0.22147454666831523 0.14071790779838228 0.0684727739511666
0.625 0.11581576345010006 0.042254303563200625 0.012116740356613078 0.0017288063394365577 0.32979170476137515 0.23801928060357452 0.15846723970175353 0.08345889628613828
0.5882352941176471 0.13321996739713987 0.054364519315024465 0.018031403465486866 0.0032528498612527 0.3408783476222913 0.2534526802153252 0.1756343159960647 0.0989402778273524
0.5555555555555556 0.15031786869103153 0.06746754961347302 0.025312180927241065 0.005567500886467251 0.3509905539833235 0.2678109103488928 0.19209874486348813 0.11463281748509924
0.5263157894736842 0.16692816530183263 0.08127762071508186 0.033868398043638545 0.00881657932217066 0.360227006991249 0.281144967837636 0.20778795029401373 0.1303029140204982
0.5 0.1829273725249763 0.09552980464799209 0.04355308018591089 0.013093577670933693 0.36867548671898637 0.29352107570239006 0.22266555145626699 0.14576513064073432
0.47619047619047616 0.19823685843113717 0.10999268152464958 0.05418628665010898 0.0184352748213388 0.3764164228709441 0.3049950214387842 0.23672217937784346 0.1608783738291687
0.45454545454545453 0.21281312362892937 0.12447243982463005 0.06557454298826298 0.024824558451135393 0.38352095826403126 0.31563558203215264 0.24996713386256264 0.17553972808319118
0.4347826086956522 0.22663889792399014 0.13881247626677157 0.07752628059675057 0.0321984496842875 0.39005272901603394 0.3255061552955211 0.26242430274274847 0.18967808643703674
0.4166666666666667 0.2397159083409105 0.1528905074610873 0.08986208704989582 0.04046116386215626 0.3960683877237928 0.3346672204894853 0.27412427430367753 0.20324794860576179
0.4 0.25205929301490043 0.16661432084258535 0.10242099009595464 0.04949483084454684 0.40161831547922994 0.3431762723099322 0.2851037097220221 0.2162238962814667
0.3846153846153846 0.26369327417426736 0.17991754398697907 0.11506339518333786 0.05917081020347247 0.40674732389138707 0.35108611213666613 0.29540218823337844 0.22859597066691634
0.37037037037037035 0.2746479320488295 0.192754352060732 0.12766974808074 0.06935839986935648 0.4114953236112051 0.35844561466871033 0.3050602191304486 0.24036582208507412
0.35714285714285715 0.28495679344790187 0.20509617322116475 0.14014926065892058 0.07993127187503887 0.4158978630658688 0.36530058949680777 0.31411810832736364 0.2515440909666336
0.3448275862068966 0.2946549202856583 0.21692782760759596 0.15241890658406682 0.09077186210224879 0.4199866383693365 0.3716922514833761 0.3226151706618175 0.2621468142170924
0.3333333333333333 0.3037782956531376 0.22824449647666323 0.16442048333211046 0.10177404081167543 0.42378994679979143 0.37765847746573944 0.33058924052153216 0.2721945292626007
0.3225806451612903 0.3123619911539783 0.2390477917586103 0.17610867678529785 0.1128443900938978 0.42733308843279344 0.38323381623122116 0.3380762940673795 0.2817104598434772
0.3125 0.3204402449417635 0.24934965309434978 0.18745058156481345 0.12390243216748273 0.4306387186685756 0.3884497452566446 0.34511025601275075 0.2907190302049834
0.30303030303030304 0.32804619985123384 0.25916191073387346 0.19842374193727425 0.1348801121496354 0.43372715635381004 0.3933349229986106 0.3517229329802023 0.2992464454974061
0.29411764705882354 0.33521054186222154 0.2685007550275516 0.20901420543873417 0.14572145139644996 0.43661664431898156 0.39791540196100794 0.35794399454556103 0.3073179364215568
0.2857142857142857 0.341963001391305 0.2773844040650694 0.21921492166322892 0.15638065652828273 0.43932360109759006 0.4022149418463873 0.36380104336105995 0.3149587138992317
0.2777777777777778 0.34833127071315545 0.28583224371622973 0.2290242366475488 0.16682153809659062 0.44186282946609345 0.40625519087433215 0.3693196957313246 0.32219334080991624
0.27027027027027023 0.3543412470705028 0.2938642766842544 0.23844488174966047 0.17701609908665109 0.44424769605191855 0.4100558491753306 0.374523655870856 0.3290454991782686
0.2631578947368421 0.36001677963164175 0.301500644360942 0.24748278912385496 0.18694342558576893 0.4464902907922483 0.41363488454969205 0.3794348512668283 0.3355378336491801
0.25641025641025644 0.36538090617338526 0.30876133235473363 0.2561464613421555 0.1965885793242986 0.4486014364623768 0.41700868640675287 0.38407353722597926 0.34169189999484234
0.25 0.3704542003934551 0.31566583857287983 0.26444619712219536 0.2059416815784823 0.4505913756999553 0.42019222031290404 0.3884584196794637 0.34752720245114993
0.24390243902439027 0.37525591605163106 0.32223323007384713 0.27239364927940324 0.21499700889173773 0.452468942879354 0.4231991659305584 0.3926067641642837 0.35306508263658387
0.23809523809523808 0.3798042313208083 0.3284817975451389 0.2800012909024191 0.22375232212431648 0.4542423450675415 0.42604203882894354 0.39653857563508127 0.35832279172222953
0.23255813953488372 0.3841156407478367 0.3344290807948103 0.28728214667495244 0.23220812077149766 0.4559190705746682 0.4287323018503178 0.40025636448979524 0.3633163985097673
0.22727272727272727 0.3882055445907799 0.3400917834771316 0.2942495050964184 0.24036715586234175 0.4575059551197252 0.431280463591462 0.40378591796975066 0.36806245671204074
0.2222222222222222 0.3920882077043041 0.3454857666689468 0.3009167133980774 0.24823377233718796 0.45900924788360914 0.4336961687166651 0.4071357155836887 0.37257589499641625
0.2173913043478261 0.39577684557984266 0.35062606589028555 0.3072970187696017 0.2558142539040327 0.46043467009885736 0.43598827829067505 0.41031735375924094 0.3768706579558663
0.2127659574468085 0.39928370957805265 0.35552688483731504 0.31340343991509756 0.2631153749845743 0.4617874660721082 0.4381649423523129 0.41334155298697456 0.38095979693356086
0.20833333333333334 0.4026201328465846 0.36020162692099245 0.3192486579079576 0.270144803758652 0.4630724505310301 0.44023366505356204 0.4162182333643837 0.3848554864933663
0.2040816326530612 0.40579663716022457 0.364662916485177 0.3248450018143286 0.2769108933744052 0.464294050029484 0.442201362969088 0.4189565771567825 0.3885691060766339
0.2 0.4088230023808471 0.3689226264597897 0.3302043477045875 0.2834222321641064 0.4654563380384701 0.444074418819441 0.4215650883318068 0.39211128226187775
0.19607843137254904 0.41170828041477214 0.37299193107181305 0.33533812893188963 0.2896876840007085 0.4665630687381061 0.4458587270654599 0.42405165463234445 0.39549193451038883
0.1923076923076923 0.41446088602988657 0.3768813325685581 0.3402572923163524 0.2957162097146144 0.4676177062720743 0.4475706497778857 0.4264235931806508 0.39872032175229666
0.18867924528301888 0.41708864018763375 0.3806006649810117 0.3449722966743362 0.30151674969098197 0.46862345089971313 0.44919368001930005 0.428687696009757 0.4018050916907795
0.18518518518518517 0.41959881540847394 0.38415919312717933 0.349493202321 0.3070981667455134 0.46958326240071097 0.4507316862392128 0.4308502740466533 0.4047543274365406
0.18181818181818182 0.4219981805840273 0.38756560451597466 0.3538294651978068 0.3124691979465256 0.47049957520369357 0.4522116291350413 0.43291719720895705 0.40757558917691733
0.17857142857142858 0.42429303216629377 0.39082804231518037 0.35799010457297864 0.31763838338419415 0.47137554439437784 0.45362635591940065 0.4348939234669877 0.4102759275834638
0.17543859649122806 0.42648923434392694 0.3939541832203261 0.36198371708318566 0.32261405467863186 0.47221322031328394 0.454979607206488 0.43678553829750066 0.4128619510540781
0.1724137931034483 0.4285922499395069 0.3969512056197032 0.3658184500918709 0.327404295663641 0.4730147931182976 0.45627486328808586 0.43859678444295863 0.4153398374080992
0.1694915254237288 0.4306072250867692 0.3998258592213263 0.3695020159255193 0.3320169076094515 0.47378229899573865 0.45751536366049317 0.4403320828537309 0.4177153721676661
0.16666666666666666 0.4325389330730974 0.4025844832103941 0.37304173602572405 0.3364594422352497 0.47451763353665516 0.4587041267210008 0.4419955619372435 0.4199939592160982
0.1639344262295082 0.434391834677934 0.40523303425706236 0.37644454808766964 0.3407391505023313 0.47522256264279306 0.45984396894409435 0.44359108060425523 0.4221806936314054
0.16129032258064516 0.43617010180733135 0.4077771116473045 0.3797170130120373 0.3448630313553615 0.47589873382066605 0.4609375187777357 0.4451222423466569 0.4242803289467254
0.15873015873015872 0.4378776390268345 0.4102219791521532 0.38286535815182 0.3488377716765014 0.4765476844055938 0.46198723188558394 0.4465924236611396 0.42629733350492804
0.15625 0.4395181042814791 0.4125725844143576 0.3858954538355008 0.35266977679981515 0.4771708511403704 0.46299518744049584 0.44800478274270744 0.42823590381567633
0.15384615384615385 0.441094929265258 0.4148335899310791 0.3888128685127094 0.35636518972324327 0.4777695774447858 0.46396398480233747 0.44936228136052464 0.4300999719315005
0.15151515151515152 0.4426113223104197 0.4170093800440156 0.39162286663618606 0.3599298798059096 0.4783451209043601 0.46489540566536597 0.4506676938863056 0.4318929863103627
0.14925373134328357 0.44407030120547714 0.4191040841855853 0.39433043591772204 0.3633694521819278 0.47889865846987933 0.46579133724212934 0.45192362540499237 0.43361893776474986
0.14705882352941177 0.4454747005082249 0.4211215951506874 0.3969402925488589 0.3666892581681859 0.4794312942060941 0.4666535500299668 0.4531325221109525 0.4352808410817518
0.14492753623188406 0.4468271922774573 0.4230655783328522 0.39945690836605346 0.36989441236903675 0.47994406467130313 0.46748370597359856 0.4542966832337263 0.4368817756249423
0.14285714285714285 0.44813026558835867 0.4249394930707564 0.40188451320385205 0.3729897766242648 0.4804379428183217 0.46828336688578787 0.4554182703627745 0.4384246438229539
0.14084507042253522 0.4493862057976353 0.42674660575711815 0.40422711517456567 0.37597999985578107 0.48091384265561654 0.4690540009883242 0.45649931773322455 0.4399121782003343
0.1388888888888889 0.45059740349390204 0.42848999650978675 0.40648850387763147 0.37886952073224567 0.4813726246314156 0.46979698963438316 0.45754174036937684 0.44134696581205085
0.136986301369863 0.4517658605951926 0.43017257585951124 0.40867227712692633 0.38166254838666697 0.4818150976356535 0.4705136336295864 0.4585473425513573 0.44273142612688265
0.13513513513513511 0.45289355922292446 0.43179710166107627 0.4107818455522071 0.38436311565083325 0.4822420237274761 0.4712051587019211 0.4595178263547489 0.44406786498909495
0.13333333333333333 0.4539823499944737 0.4333661579997764 0.41282043955709624 0.3869750524778983 0.48265412108334727 0.4718727201270185 0.4604547971672306 0.44535844989508827
0.13157894736842105 0.4550339969039042 0.43488220572501196 0.41479112436498616 0.3895020057003093 0.48305206702074893 0.47251740788734625 0.46135976901482967 0.4466052318192925
0.12987012987012986 0.45605015412251404 0.43634758465405193 0.4166968074530492 0.3919474706823298 0.48343650068904853 0.4731402506473687 0.46223417416112417 0.44781014749948006
0.12820512820512822 0.4570323949965595 0.4377644917367089 0.41854025011243084 0.3943147582426633 0.48380802572351345 0.47374222078608 0.4630793666534854 0.44897503494167856
0.12658227848101264 0.4579822038578836 0.43913501410252476 0.4203240696002305 0.39660703142702497 0.4841672123963781 0.47432423677120356 0.46389662385470204 0.4501016215935803
0.125 0.4589009792305182 0.44046112896128636 0.42205075532977154 0.39882730427357993 0.4845145999842208 0.47488716814655096 0.4646871572679697 0.4511915645345996
0.1234567901234568 0.45979006099333536 0.4417447135485887 0.4237226696320313 0.4009784509619493 0.48485069867870895 0.47543183765199637 0.46545211481519877 0.4522464059045144
0.12195121951219513 0.4606507048684185 0.4429875462831624 0.42534205978860795 0.4030632124503504 0.48517599142936646 0.47595902389778477 0.4661925796464816 0.4532676356861267
0.12048192771084336 0.4614841043137459 0.44419131558544067 0.4269110613556085 0.40508419610744256 0.48549093564437873 0.47646946501753235 0.46690958261198917 0.45425665172037305
0.11904761904761904 0.46229139219581267 0.44535762445841226 0.4284317050001587 0.4070439036171947 0.48579596492680105 0.47696386049963935 0.4676040990293349 0.4552147817178576
0.11764705882352941 0.46307364320260225 0.44648799635337433 0.42990592441590125 0.4089447023351759 0.4860914903933792 0.477442873943293 0.4682770554295024 0.45614328881546373
0.11627906976744186 0.46383187734967996 0.447583878608492 0.43133555713144023 0.4107888689501768 0.48637790210174475 0.47790713508747745 0.46892933164328543 0.4570433745203926
0.1149425287356322 0.4645670639603362 0.4486466469705881 0.4327223566386345 0.41257856049122227 0.48665557039037716 0.47835724207123276 0.4695617622091103 0.45791617902766923
0.11363636363636363 0.46528012412398073 0.44967761120555516 0.43406725659271644 0.41431585114916825 0.48692484700031485 0.47879376269591495 0.4701751396088437 0.4587627818224443
0.11235955056179775 0.4659719330470226 0.4506780169102061 0.435373308930482 0.41600271279732065 0.4871860661447457 0.4792172370135984 0.47077022022030657 0.4595842186769156
0.1111111111111111 0.46664332433214195 0.45164905026863 0.43664129947229136 0.4176410304545136 0.4874395456173243 0.47962817835135124 0.47134772140314635 0.4603814671303397
0.10989010989010989 0.4672950905987501 0.4525918442717612 0.4378726731354401 0.4192326023399392 0.48768558773200316 0.48002707543699963 0.4719083247842776 0.46115546546024994
0.10869565217391305 0.46792798710420036 0.45350746782672624 0.4390688085749547 0.42077915961388135 0.48792448020691737 0.48041439327508617 0.4724526823099937 0.46190710251772477
0.1075268817204301 0.4685427320672958 0.4543969509968704 0.4402310199501534 0.4222823356247867 0.48815649700366326 0.48079057449669554 0.47298141336001054 0.4626372238668619
0.10638297872340426 0.469140011003114 0.45526127469228583 0.44136056298259896 0.4237437081999057 0.4883818990994541 0.4811560407825176 0.47349510770627423 0.4633466419165733
0.10526315789473684 0.4697204780163093 0.4561013712409271 0.44245863668068 0.4251647837399187 0.4886009351987147 0.48151119502151013 0.4739943282555115 0.4640361252021744
0.10416666666666667 0.4702847556131221 0.4569181317521199 0.4435263852417277 0.426547004701995 0.4888138425222709 0.481856420361783 0.4744796127727262 0.46470640932701307
0.10309278350515465 0.4708334389944193 0.4577124072399024 0.44456490776644547 0.4278917425801574 0.48902084728177847 0.4821920829277395 0.474951472646861 0.46535820137219996
0.1020408163265306 0.47136709571153806 0.4584850103830221 0.44557523921756814 0.429200316369918 0.4892221654212028 0.4825185315714491 0.4754103971299362 0.4659921666851586
0.10101010101010101 0.47188626814058354 0.4592367183763777 0.4465583904645556 0.4304739905821732 0.48941800302133637 0.482836100402059 0.4758568534933591 0.46660894787148177
0.1 0.47239147438688006 0.4599682734596366 0.44751531605501 0.43171397724175653 0.4896085570084992 0.4831451073566785 0.4762912885422523 0.4672091570565996
}\dataAccuracyOverSigmaReplica